\documentclass[twocolumn,preprintnumbers,amsmath,amssymb,superscriptaddress,nofootinbib,longbibliography,floatfix,prx]{revtex4-2}
\usepackage[utf8]{inputenc}
\usepackage[english]{babel}

\usepackage{graphicx}

\usepackage{leftidx}
\usepackage{diffcoeff}
\usepackage{amsfonts,amssymb,amsmath}
\usepackage{mathtools}
\usepackage{bm}
\usepackage{bbm}
\usepackage{units}
\usepackage{mathrsfs}
\usepackage{upgreek}
\usepackage{array} 
\usepackage{tabularx}
\newcolumntype{Y}{>{\centering\arraybackslash}X}
\usepackage{multirow}
\usepackage{xcolor}

\newcommand{\ee}[1]{{\rm e}^{{#1}}}

\newcommand{\rd}{{\rm d}}
\newcommand{\veps}{\varepsilon}
\newcommand{\bchi}{\boldsymbol{\chi}}
\newcommand{\bp}{\bar{p}}
\newcommand{\bv}{\bar{\textbf{v}}}

\colorlet{mylinkcolor}{blue!66!black!80}
\usepackage[colorlinks=true,linkcolor=mylinkcolor,citecolor=mylinkcolor,filecolor=cyan,urlcolor=mylinkcolor,breaklinks=true]{hyperref}

\begin{document}

\title{Multiple Pareto-optimal solutions of the dissipation-adaptation
  trade-off}
\date{\today}

\author{Jorge Tabanera-Bravo}
\email{jtabane@mpinat.mpg.de}
\author{Alja\v{z} Godec}
\email{agodec@mpinat.mpg.de}
\affiliation{Mathematical bioPhysics group, Max Planck Institute for Multidisciplinary Sciences, Göttingen 37077, Germany}

\begin{abstract}
Adaptation refers to the ability to recover and maintain ``normal''
function upon perturbations of internal \textcolor{black}{or} external
conditions and is essential for sustaining life. Biological  adaptation mechanisms are dissipative, i.e.\ they require a supply of energy such as the coupling to
the hydrolysis of ATP. Via evolution the underlying biochemical machinery of
living organisms evolved into highly optimized states. However, in the
case of adaptation processes two quantities are optimized simultaneously, the
adaptation speed or accuracy and the thermodynamic cost. In such
cases one typically faces a trade-off, where
improving one quantity implies worsening the other. The solution is no
longer unique but rather a \emph{Pareto set}---the set of all physically
attainable protocols along which no quantity can be improved without
worsening  another. Here we investigate  Pareto
fronts in adaptation-dissipation
trade-offs for a cellular thermostat and a minimal ATP-driven
receptor-ligand reaction network. We find convex sections of Pareto
fronts to be interrupted by concave regions, implying the coexistence
of distinct optimization mechanisms. We discuss the implications of
such ``compromise-optimal'' solutions and argue that they
may endow biological systems with a superior flexibility
to evolve, resist, and adapt to different environments. 
\end{abstract}

\maketitle

\section{Introduction}
Living organisms and in particular their sub-units  
operate in contact with complex noisy environments, constantly exchanging energy
and matter. Thereby far-from-equilibrium conditions are not only
non-detrimental but are in fact essential for
proper biological
function \cite{Astumian2002Nov, Astumian2019Aug, Amano2022May}. The
biochemical processes involved are able to remain remarkably precise, stable,  and
responsive even at high noise levels. A
paradigmatic example of such processes is ``kinetic proofreading'' in gene transcription
\cite{Boeger2022Jun, Qian2007May}, where the cell exploits ATP
hydrolysis to improve the precision of gene transcription at the cost
of increasing energy dissipation. Another perquisite for sustaining
intact biological activity (and ultimately life) is \emph{adaptation}---the ability
to self-regulate upon a perturbation of internal or external
conditions in order to recover and maintain normal
function \cite{Ma2009Aug}. Examples include the adaptation
of sensory systems to different light conditions
\cite{Smirnakis1997Mar}, chemical gradients \cite{Hazelbauer2008Jan}
or temperature shocks \cite{Thieringer1998Jan, Etchegaray1996Feb,
  Buckley2001Oct}. A biological system may be able to activate
drastically different mechanisms to adapt to 
changes in the
environment. For example, under heat- \cite{Sriram2012Aug,
  Sivery2016Dec, Buckley2001Oct} or cold-shocks
\cite{Etchegaray1996Feb} a system expresses different genes
depending on
temperature variations.
\\
\indent 
By evolutionary arguments it is conceivable that adaptation mechanisms in Nature are optimized to maximize accuracy. Such optimization could
explain this sensitive 
switching of adaptation mechanism under small variations of
conditions. A deeper discussion requires a precise definition of the
parameters and observables to be optimized. 
\\
\indent 
Optimal protocols in stochastic systems have already 
been addressed in the recent literature \cite{Schmiedl2007Mar,
  Falasco2022Feb, Guery-Odelin2023Jan}. Typical problems correspond
to finding the protocol extremizing a certain thermodynamic quantity, such as the mean work \cite{Schmiedl2007Mar} or the time
\cite{Plata2020Mar} required to drive a system between two different
states.
\\
\indent 
A more general problem involves the simultaneous optimization of two or more
quantities. In most cases, one cannot optimize  these
quantities simultaneously. Instead one faces a trade-off, where
improving one quantity implies worsening the others, for example, the
average work required to drive a system between two states versus the
work fluctuations \cite{Horowitz}. The solution of such optimization problems is no
longer a unique protocol, but the set containing all physically
attainable protocols along which no quantity can be improved without
worsening  another \cite{Censor1977Mar}. Such sets are known as \emph{Pareto
fronts} and are in fact ubiquitous in Nature, beyond those encountered
in stochastic
thermodynamics \cite{Shoval2012Apr, Seoane2016May, Dinis2020Nov}. Each protocol in
the Pareto front is called \emph{Pareto optimal}. 
\\
\indent 
The trade-off between precision and dissipation has already been
discussed in the context of thermodynamic uncertainty relations
\cite{Barato2015Apr,Barato2015Jun,Seifert2019Mar,England,Hororwitz_2019,Udo_a,Hartich,Dieball}
and Pareto fronts for non-autonomous systems \cite{Proesmans}.
However, there seem to have been only a few attempts to relate
dissipation to adaptation in complex systems \cite{Lan2012May}.
In particular, the 
dissipation-adaptation trade-off 
has 
\emph{not} yet been systematically addressed in terms
of Pareto optimality. 
\\
\indent Here we investigate Pareto fronts in the simultaneous optimization of
dissipation and adaptation in 
minimal autonomous biomolecular systems---a \emph{cellular thermostat}
and a \emph{receptor-ligand reaction network}---and describe
their ability to adapt to different kinds of environmental changes.
We determine optimal configurations which maximize
adaptation accuracy under minimum entropy production and discuss their
physical implications. In particular, we consider the adaptation of the
cellular thermostat---a HSP/CPS 
 protein expression network \cite{Thieringer1998Jan,
   Etchegaray1996Feb, Buckley2001Oct, Sriram2012Aug,
   Sivery2016Dec}---to abrupt and smooth, time-periodic
 (i.e.\ circadian \cite{Mehra2006Jul, dobroborsky2006thermodynamics})
 temperature changes, as well as the adaptation-dissipation trade-off
 in \color{black} a receptor-ligand chemical network, a fundamental element of kinetic proofreading \cite{Boeger2022Jun, Qian2007May} and cellular response mechanisms \cite{Qian2005Jan, Goldbeter1981Nov, Koshland1982Jul}. Both
 systems are shown to display multiple Pareto-optimal solutions and we
 discuss their biophysical implications. \color{black}

\subsection{Structure of the manuscript}

In Secs.~\ref{subsec:feedback_model} and \ref{subsec:feedback_1} we recapitulate a model of the action of HSP
proteins upon abrupt temperature changes \cite{Thieringer1998Jan,
  Etchegaray1996Feb,Buckley2001Oct, Sriram2012Aug, Sivery2016Dec}, 
present results for a single shock, and discuss the trade-off between adaptation
accuracy and dissipation. 
In Sec.~\ref{subsec:feedback_2} we consider
time-periodic shocks and harmonic temperature changes, representing, for example,
the daily temperature variation driving the circadian response
\cite{Mehra2006Jul, dobroborsky2006thermodynamics}. We show that such multiple and
continuous temperature changes drastically alter the
adaptation-dissipation trade-off, in particular
the structure of optimal solution sets. Finally, in
Sec.~\ref{subsec:feedback_3} we consider the recovery from random abrupt temperature
shocks, and explain why the shock intensity determines whether or not
the system can efficiently adapt to such perturbations. 
In Sec.~\ref{sec:chemical} we address \color{black}adaptation \color{black} in a receptor-ligand reaction network (involved e.g.\ in protein synthesis)
and study its adaptation
to changes of ligand concentration. We determine
optimal solutions of the adaptation-dissipation trade-off and discuss
the mechanisms underlying the different optimal solutions. \color{black}We conclude with a hypothesis
on why optimal solutions distributed along extended fronts may be
biologically beneficial, and compare with recent experimental results.  
\color{black}

\section{Cellular thermostat}\label{sec:feedback}

Living systems responded to environmental perturbations in several ways
\cite{pigllucci1996organisms}. Often this response involves changes in
gene expressions. For example, under temperature shocks
\cite{Buckley2001Oct, Etchegaray1996Feb, Thieringer1998Jan} some
organisms respond by producing so-called heat-shock proteins (HSP) or cold-shock
proteins (CSP) depending on the respective sign of the temperature
change. A heat shock may cause some proteins to become denaturated
\cite{Buckley2001Oct}, whereas a cold shock can inhibit protein
synthesis \cite{Thieringer1998Jan}. During heat shocks HSP proteins act as molecular chaperones
that stabilize proteins \cite{Marchler2001Feb, Feldman2000Feb}. Conversely,
during a cold shock, an attenuation of protein synthesis becomes
counteracted by CSP \cite{Thieringer1998Jan}. 

\begin{figure*}[htb!]
    \centering
    \includegraphics[width = \linewidth]{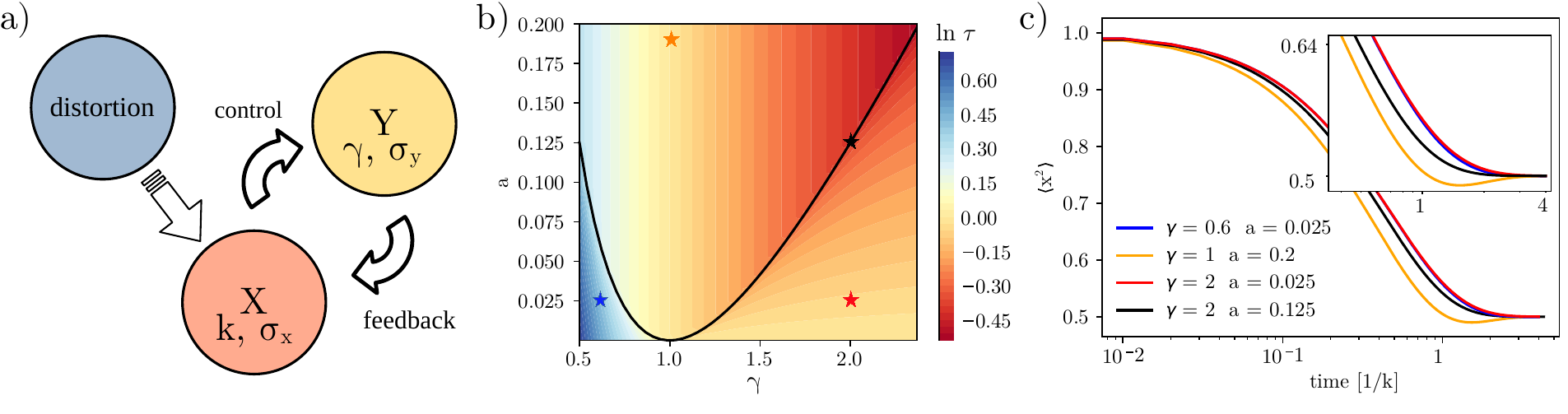}
    \caption{(a)~Schematic of feedback adaptation mechanism between
      $x_t$ and $y_t$.~(b) Relaxation timescale $\tau$ on a
        logarithmic scale as a function of $a$ and $\gamma$. The black
      line denotes $a(\gamma) = (k-\gamma)^2 / 4\gamma$. The stars
      highlights the points considered in panel (c). (c) Evolution of
      $\langle x_t^2 \rangle$ for several values of $\gamma$ and $a$
      (see starts in (b)). The initial state was chosen as $\langle
      x_0^2\rangle = \langle y^2_0\rangle = 1$ and $\langle
      x_0y_0\rangle  = -0.1$ and we set $k = 1$, $\sigma_x = 1$ and $\sigma_y = \sigma_x / \sqrt{ka}$. The inset shows a magnification of the
      large-time region.}
    \label{fig:Eigenvalues}
\end{figure*}

We consider a minimal model for the concentration
(more precisely activity) of some functionally relevant protein at
time $t$, denoted by $x_t$, after a temperature shock. The shock triggers the
induction of HSP, whose concentration is denoted by $y_t$, and acts to
restore the concentration $x_t$ to its ``normal'' state. This feedback action is sketched in Fig.~\ref{fig:Eigenvalues}a;
if the activity $x_t$ deviates from its ``normal'' value, the control
$y_t$ is
activated and in turn acts to dampen the fluctuation of $x_t$, forcing
it back to its ``normal'' value. The molecules $x$ and $y$
are coupled to a common Gaussian (white-noise) thermal bath at a given
temperature but with different couplings, leading to noise intensities 
$\sigma_x$ and $\sigma_y$, respectively. The concentrations $x_t,y_t$  evolve according to the (It\^o)  Langevin equations,
\begin{equation}\label{eq:langevin}
   \left(\begin{matrix}
        \rd x_t\\
        \rd y_t
    \end{matrix}\right)
    \!=\!
    \left(
    \begin{matrix}
        -k & -a \\
        \gamma& -\gamma\\
    \end{matrix}
    \right) \left(\begin{matrix}
         x_t\\
         y_t
    \end{matrix}\right)\rd t
    +\left(
    \begin{matrix}
        \sigma_x & 0 \\
        0 & \gamma\sigma_y\\
    \end{matrix}
    \right) 
     \left(\begin{matrix}
         \rd W^x_t\\
         \rd W^y_t
    \end{matrix}\right),
\end{equation}
where $k$ represents the restitution constant of $x_t$, $a$ the
back-action of $y_t$ on $x_t$, and $\rd W^{x,y}_t$ are increments of
the Wiener process obeying $\langle \rd W^{x,y}_t\rd W^{x,y}_{t'}\rangle=\delta_{x,y}\delta(t -
t')\rd t\rd t'$. The parameter $\gamma^{-1}$ sets the time-scale on which $y_t$ responds to changes in $x_t$ \cite{Munakata2013Jun, Munakata2014May, Horowitz2014Dec}. A similar model has been employed in \cite{Lan2012May} to describe the
relation between adaptation and energy consumption in \textit{E. coli}
chemotaxis \cite{Tu2008Sep} and a discrete version of the model was
investigated in \cite{Sartori2014Dec}. In both cases, the feedback
tends to rectify the average value $\langle x_t\rangle$ towards the
equilibrium under the action of an external signal. However, the same
adaptation mechanism can be used to control, not only the
average but also the dispersion (i.e. the variance) \cite{Conti2022Nov}. Since we are
interested in describing a thermostat, we will adapt the dispersion.
We can set, without any loss of generality, the average to zero and
define the variance as $\langle x_t^2\rangle$. 

We consider zero-mean Gaussian initial conditions, and because
the evolution equation \eqref{eq:langevin} is an Ornstein-Uhlenbeck process, the probability density
remains Gaussian for all times and
is fully characterized in terms of the (symmetric) covariance matrix with
elements $\langle x^2_t\rangle,\langle x_ty_t\rangle,\langle
y^2_t\rangle$ which for notational convenience we write in a
vector $\textbf{v}_t = (\langle x_t^2\rangle, \langle x_ty_t\rangle, \langle y_t^2\rangle)$. From Eq.~\eqref{eq:langevin} follows the Lyapunov equation
\begin{equation}
    \dot{\textbf{v}}_t = M\textbf{v}_t + \textbf{d}
    \label{eq:Lyapunov}
\end{equation}
where we introduced the vector $\textbf{d} \equiv (\sigma_x^2, 0,
\gamma^2\sigma_y^2)^T$ and 
\begin{equation}
    M = \left(
    \begin{matrix}
        -2k & -2a & 0\\
        \gamma& -\gamma - k & -a\\
        0 & 2\gamma & -2\gamma
    \end{matrix}
    \right).
 \label{MM}   
\end{equation}
The solution of Eq.~\eqref{eq:Lyapunov} reads
\begin{equation}
    \textbf{v}_t = \left( 1 - \ee{Mt}\right)\textbf{v}_{\rm s} + e^{Mt}\textbf{v}_0,\label{eq:state_on_time}
\end{equation}
which has the steady-state solution $\textbf{v}_{\rm s} = -M^{-1}\textbf{d}$. 

The efficiency of $y_t$ to control $x_t$ is encoded in the positive-definite drift matrix in
Eq.~\eqref{eq:langevin}, i.e.\
\begin{equation} 
    A = \left(
    \begin{matrix}
        k & a \\
        -\gamma& \gamma\\
    \end{matrix}
    \right),
\label{AA}    
\end{equation}
which has  eigenvalues
\begin{equation}
    \lambda_\pm = (k+\gamma)/2 \pm \sqrt{(k-\gamma)^2/4 - a\gamma}.
    \label{eq:eigenvalues_expression}
\end{equation}
Large real eigenvalues imply rapid control of fluctuations of $x_t$.
The relaxation timescale of the controlled system upon a strong fluctuation
is $\tau = \min[\rm{Re}(\lambda_+),\rm{Re}(\lambda_-)]^{-1}$.
The relaxation time $\tau$ is shown 
in Fig.~\ref{fig:Eigenvalues}b as a function of $a$ and $\gamma$. In
the non-interacting scenario  $a = 0$ the subsystem $x_t$ relaxes on
the timescale $1/k$ and $y_t$ on the timescale $\tau = 2/(k + \gamma)
= 1$. For $\gamma < k$ the timescale increases (blue region in
Fig.~\ref{fig:Eigenvalues}b) and here the feedback loop is
\emph{not} able to accelerate the relaxation. Only for high values of $a$ and $\gamma$
we observe a significant relaxation acceleration (red region in
Fig.~\ref{fig:Eigenvalues}b).
An important set of $a,\gamma$ is the line $(k-\gamma)^2 - 4 a\gamma = 0$
(Fig.~\ref{fig:Eigenvalues}b, black curve). Above this line, the eigenvalues in Eq.~\eqref{eq:eigenvalues_expression} become
complex, which may create oscillatory transients with frequency $\omega = [a\gamma -
(k-\gamma)^2]^{1/2}/2$  during the relaxation
of the whole system. The time evolution of the dispersion of the controlled species
$x_t$,  $\langle x_t^2\rangle$, is shown in
Fig.~\ref{fig:Eigenvalues}c for several values of $\gamma$ and $a$
highlighted in Fig.~\ref{fig:Eigenvalues}b as stars.  For $\gamma =
2$, $a = 0.025$ as well as  $\gamma = 0.6$, $a = 0.025$ the relaxation
rates are similar, since $x_t$ is essentially freely
relaxing. Conversely, for $\gamma = 1$, $a = 0.2$ the relaxation
occurs in the region where eigenvalues are complex, with relaxation
time $\tau = 1$ and frequency $\omega = \sqrt{0.2} \approx 0.44$, the
oscillations are hence strongly damped (see inset in
Fig.~\ref{fig:Eigenvalues}c).  The pair $\gamma = 2$, $a = 0.125$
is taken on the black curve and displays a faster relaxation and
without any oscillations.

\subsection{Adaptation of the cellular thermostat}\label{subsec:feedback_model}

It is a common theme in biology that systems are able to remain
in (or close to) homeostasis despite persistent perturbations
\cite{Ma2009Aug, Lan2012May}, a process referred to as
\textit{adaptation}. In the case of the cellular
thermostat adaptation refers to determining the region of
parameters $\gamma, a$ and $\sigma_y$, for which $x_t$ evolves exactly into the desired homeostatic state $\langle
x^2\rangle_{\rm s} = \sigma_x^2/2k$. Once this region has been
determined, any perturbation can be suppressed by increasing $\alpha$ and
$\gamma$ along the restricted values. 
The region of adapted parameters can be determined from Eq.~\eqref{eq:Lyapunov}, by solving the set of algebraic equations $\textbf{v}_{\rm s} = (\sigma_x^2/2k, 0, \langle y^2\rangle_{\rm s})^T = -M^{-1}\textbf{d}$ for $\gamma, a$ and $\sigma_y$. Note that we are imposing \emph{no} constraints on the final distribution of $y_{t\to\infty}$, so the system is not completely determined. We address this problem in Appendix~\ref{app:steady}, where we prove that the constraint
\begin{equation}
  \sigma_y^2 = \frac{\sigma_x^2}{ka}
  \label{eq:condition}
\end{equation}
implies the required adaptation $\langle x^2\rangle_{\rm s} =
\sigma_x^2/2k$. In this case, both $\gamma$ and $a$ remain
undetermined, and $\langle y^2 \rangle_{\rm s} = \gamma \sigma_y^2/2
= \gamma \sigma_x^2/2ka$. As an illustrative example we consider the
condition Eq.~\eqref{eq:condition} in the relaxation processes in
Fig.~\ref{fig:Eigenvalues}c and confirm that different evolution
parameters indeed yield the same final state. \textcolor{black}{For
simplicity, we consider $\sigma_y$ to be an adjustable parameter. In
experiments this may be achieved by varying parameters, such as $a$,
the environmental temperature, and via some directed mutation perhaps the response timescale $\gamma^{-1}$.}

We quantify the velocity of adaptation process in terms of the \textit{adaptation error}
\begin{equation}
    \epsilon(a, \gamma, t) = |\langle x_t^2\rangle - \langle
    x^2\rangle_{\rm s}|,
    \label{eq:adaptation_error_definition}
\end{equation}
which quantifies how far $x_t$ from $\langle x^2\rangle_{\rm s}$ at a given time $t$. Using Eq.~\eqref{eq:state_on_time} we can rewrite $\epsilon$ as
\begin{equation}
    \epsilon(a, \gamma, t) = |\textbf{e}^T_0\ee{Mt}(\textbf{v}_0 -
    \textbf{v}_{\rm s})|,
    \label{eq:adaptation_error}
\end{equation}
with $\textbf{e}_0 = (1,0,0)^T$. It follows that $\epsilon(a, \gamma,
t)\rightarrow 0$ when $t \rightarrow 0$ independent of $\alpha$ and
$\gamma$. Along the same lines $\varepsilon(a, \gamma, 0)$ is only
given by our choice of $\langle x_0^2 \rangle$. In any intermediate
state, the adaptation error depends on  $\alpha$ and $\gamma$, as well
as on $\langle y_0^2 \rangle$. 

The adaptation error $\epsilon$ is shown in
Fig.~\ref{fig:Pareto_fronts}a  as a
function of $\gamma$ for several values  $a$ at fixed
time $t = k = 1$. According to Fig.~\ref{fig:Eigenvalues}c $x_t$ 
at this time is \emph{not} yet completely relaxed. 
The points represent different values of $a$ and $\gamma$, the blue points
satisfy $a < (k-\gamma)^2/4k$ and therefore arise in the region with
no oscillations (see Fig.~\ref{fig:Eigenvalues}). The orange points
satisfy $a > (k-\gamma)^2/4k$ and lie in the non-monotonic relaxation
regime. $\epsilon$ remains high for any $a$ and low values of
$\gamma$, since there the control $y_t$ reacts too slowly to changes of
$x_t$. Increasing $\gamma$ we improve the control of $x_t$, reducing
$\epsilon$. The black line coincides with the black line in
Fig.~\ref{fig:Eigenvalues}b and corresponds to $a =
(k-\gamma)^2/4\gamma$, while the red line contains the points with
minimal adaptation error $\epsilon$ for every $\gamma$. The black and red curves
coincide for high values of $\gamma$ and differ for low values. 

In order to explain this observation, we must take into account the
adaptation condition Eq.~\eqref{eq:condition}. According to
Eq.~\eqref{eq:condition} the final distribution of $y_t$ depends
on $a$. Therefore, if $\langle y^2\rangle_{\rm s}(\gamma, a)$ moves
far from $\langle y_0^2\rangle$, the complete system requires
additional time to adapt. This fact is especially relevant for low
values of $a$. We can represent this result by defining 
\begin{equation}
  a = \frac{\gamma \sigma_x^2}{2k\langle y^2\rangle_{\rm 0}},
  \label{eq:curva_verde}
\end{equation}
which contains all the points satisfying $\langle y^2\rangle_{\rm
  eq}(\gamma, a) = \langle y_0^2\rangle$. Along this line the system $y_t$
requires \emph{no} extra time to adapt rendering relaxation efficient (see green line in
Fig.~\ref{fig:Pareto_fronts} and note that it coincides with the red line for low values of $\gamma$).

\subsection{Adaptation-dissipation trade-off} \label{subsec:feedback_1}

An important consequence of the adaptation
condition~\eqref{eq:condition} is that it leads to a non-equilibrium
situation where detailed balance is broken. It implies a entropy production,
even in the steady state. Since $\sigma_y$ (as parameterized) depends
on $a$, decreasing the adaptation error implies an increasing entropy
production. 
The total entropy production can be separated into (non-negative)
\textit{adiabatic} and \textit{non-adiabatic}
contributions~\cite{VandenBroeck2010Jul}, the former being defined as
\begin{equation}
    \dot S_{\rm a}(t) = \int  p_t(\textbf{x})\left(\frac{\textbf{J}_{\rm s}(\textbf{x})}{p_{\rm s}(\textbf{x})}\right)^T\!D^{-1}\!\left(\frac{\textbf{J}_{\rm s}(\textbf{x})}{p_{\rm s}(\textbf{x})}\right)d\textbf{x}\geq 0, \label{eq:adiabatic_ent_definition}
\end{equation}
where $\textbf{x} = (x,y)$, $p_t(\textbf{x})$ is the joint probability
distribution of $x_t$ and $y_t$, $p_{\rm s}(\textbf{x})\equiv p_{t\to\infty}(\textbf{x})$ is the
steady-state probability density. If we define the positive-definite
diagonal diffusion matrix $D\equiv{\rm diag}\,( \sigma_x^2/2,\gamma^2\sigma_y^2/2)$
then the steady-state (i.e.\ invariant) current of the system in Eq.~\eqref{eq:langevin}
 is given by $\textbf{J}_{\rm
  s}(x,y) = -A\textbf{x}p_{\rm s}(\textbf{x}) - D\nabla
p_{\rm s}(\textbf{x})$. Note that the adiabatic entropy production
corresponds to the total entropy production in a non-equilibrium
steady state. 

The non-adiabatic entropy production represents the additional
dissipation due to the relaxation during transient \cite{VandenBroeck2010Jul,Diebal_asymm,Ibanez_2024,Cai_transport}. Even if both
contributions are non-zero during the relaxation, we will focus on the
adiabatic contribution alone. The reason for this choice is that $S_{\rm a}$
encodes thermodynamic forces
breaking time-reversal symmetry, that continuously
``steer'' relaxation to the non-equilibrium rather than the equilibrium
steady state. In addition, assuming
Gaussian initial distributions
Eq.~\eqref{eq:condition}, we can rewrite Eq.~\eqref{eq:adiabatic_ent_definition} in a simple way in terms of
$\langle x_t^2\rangle$ and $\langle y_t^2\rangle$ (see
Appendix~\ref{sec:adiabatic_appendix}) as
\begin{equation}
    \dot S_{\rm a}(t) = \frac{2a}{\sigma_x^2}(a\langle y_t^2\rangle + k\langle x_t^2\rangle) = \textbf{s}^T\textbf{v}_t.
    \label{eq:adiabatic_entropy}
\end{equation}
where $\textbf{s} = [2a/\sigma_x^2(k,0,a)]^T$, and $\textbf{v}_t$ is
given by Eq.~\eqref{eq:state_on_time}. In turn, the adiabatic entropy
produced until time $t$ reads
\begin{align}
    \Delta S^t_{\rm a} &= \int_0^t \dot S_{\rm a}(\tau) d\tau \nonumber
    \\
    & = \textbf{s}^T\,\left[ \textbf{v}_{\rm s}t + M^{-1}\left(\ee{Mt} - 1 \right)(\textbf{v}_0-\textbf{v}_{\rm s})\right].     
\label{split}
\end{align}
The first term represents the entropy produced in the steady state in
the interval $[0,t]$, given by $\textbf{s}^T\, \textbf{v}_{\rm s}t =
a(1 + \gamma/k)t$ which grows linearly with the control parameters
$\alpha$ and $\gamma$. The second term represents the additional
adiabatic entropy produced during the transient which can be positive or negative but is always larger than $-\textbf{v}_{\rm s}t$. The splitting
in Eq.~\eqref{split} suggests that the thermodynamic cost of the
transient is essentially the cost of maintaining the non-equilibrium
steady state (as if it were already reached) augmented by the
contribution of the transient that depends on the initial condition.

\begin{figure*}
    \centering
    \includegraphics[scale = .5]{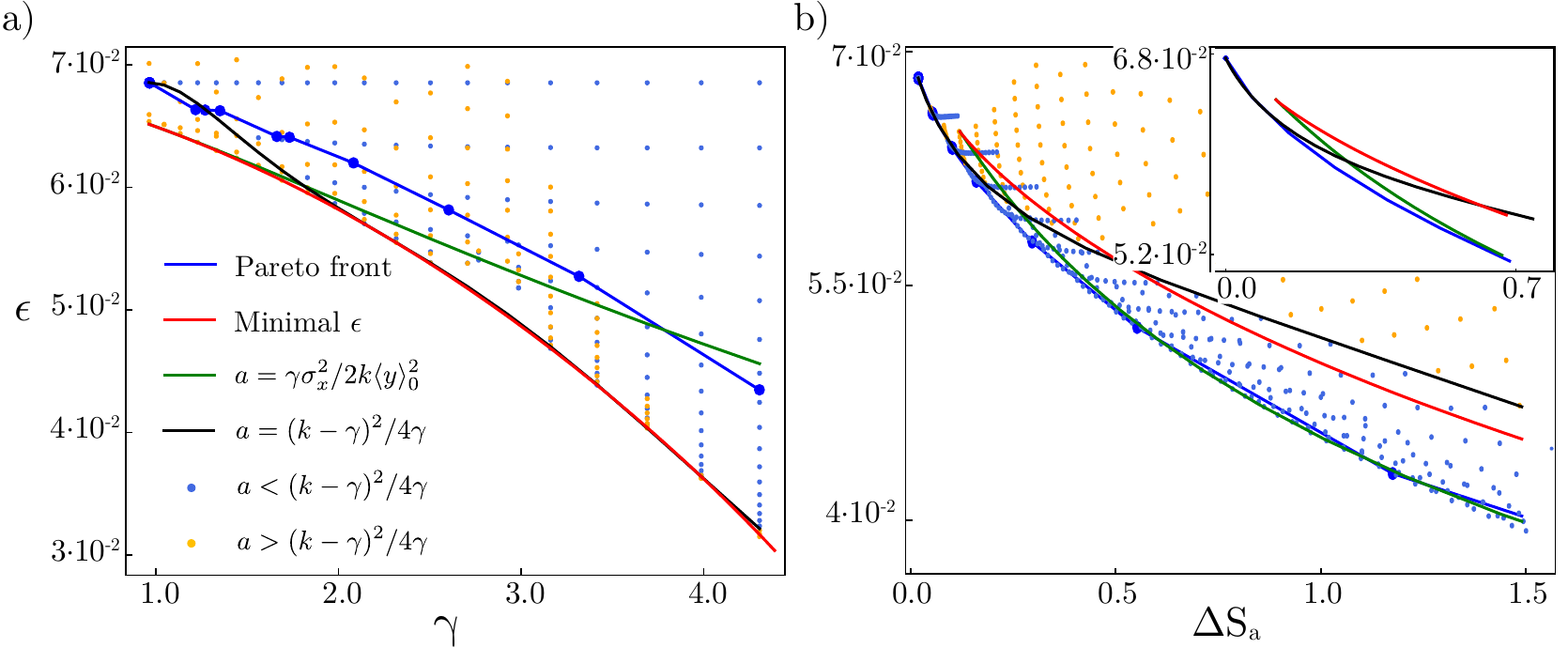}
    \caption{(a)~Adaptation error as a function of $\gamma$ for
      several values of $a\in [10^{-3},2]$ at time $t = 1$. Each 
      point represent a different value of $\gamma$ and $a$.
      Point in orange lie in the oscillatory region, $a >
      (k-\gamma)^2/4k$,
      the red line represents the result obtained by 
      numerical gradient descent and the blue line is the Pareto front. The
      black line is given by $a(\gamma) = (k-\gamma)^2/4\gamma$ and
      the green one by $a = \gamma \sigma_x^2 / 2k\langle
      y_0^2\rangle$. (b)~Error rate as a function of the
      adiabatic entropy change $\Delta S_{\rm a}^t$ for the same values
      of $a$ and $\gamma \in [1,6]$. The color code is the same as in
      panel (a).~Inset:~Magnification of the top left
      region. The initial state was chosen to be $\langle
      x_0^2\rangle = \langle y_0^2\rangle = 10$ and $\langle
      x_0y_0\rangle  = -0.1$. In both panels $k = 1$, $\sigma_x = 1$ and
      $\sigma_y = \sigma_x / \sqrt{ka}$.} 
    \label{fig:Pareto_fronts}
\end{figure*}

In Fig.~\ref{fig:Pareto_fronts}b we represent $\epsilon$ as a function
of $\Delta S_{\rm a}^t$ at time $t = 1$, for the values of $a$ and $\gamma$
from Fig.~\ref{fig:Pareto_fronts}a.
We observe that all points lie in a region with a convex
boundary. This means that
we can adjust the parameters $a$ and $\gamma$ to reduce $\epsilon$
until we reach the convex boundary. At this point, decreasing
$\epsilon$ always implies an increasing entropy production and
\textit{vice versa}. This boundary is known as the \textit{Pareto front},
and all the points along the front are called \textit{Pareto optimal}. 

A simple method to determine the Pareto front \cite{Seoane2016May} is via
the function
\begin{equation}
    J(a,\gamma;\alpha) = \alpha\epsilon(a, \gamma, t) +
    (1-\alpha)\Delta S_{\rm a}(a, \gamma, t)
    \label{conv} 
\end{equation}
where $\alpha \in (0,1)$. The Pareto front $\mathcal{P}$ then corresponds to the
set of points $(\gamma, a)(\alpha)$ that minimize $J$ for each
$\alpha$
\begin{equation}
  \mathcal{P}\equiv\inf_{\gamma,
      a|\alpha}J(a,\gamma;\alpha)\quad \forall \alpha \in (0,1).
  \label{inf}
\end{equation}  
The only requirement for this method to work is that the Pareto front
is convex. The presence of a concave region in the Pareto front in the $(\epsilon,
\Delta S_{\rm a})$ space in turn implies a sharp jump of the front in the
$(\gamma, a)$ space. This fact has been investigated in detail in
\cite{Seoane2016May, Horowitz} and will be relevant in the
following sections.  

In Fig.~\ref{fig:Pareto_fronts}b we determine the Pareto front
for the cellular thermostat
numerically (blue line). It is completely
convex and does \emph{not} coincide with the red line, since the maximal
reduction of $\epsilon$ implies a larger entropy production, pushing
the red line away from the front to the right (see Fig.~\ref{fig:Pareto_fronts}b). However, we observe that the
structure of the Pareto front is related to the solution sets
$a(\gamma) = (k-\gamma)^2/4\gamma$ (black line) and $a =\gamma
\sigma_x^2/2k\langle y^2\rangle_{\rm 0}$ (green line, Eq.~\eqref{eq:curva_verde}); see inset of Fig.~\ref{fig:Pareto_fronts}b. For low $\Delta
S_{\rm a}$, or weak control of $x$, the Pareto front is close to the
black line where irreversible driving is more efficient. Increasing $\Delta
S_{\rm a}$ the front moves away from the black line and approaches the
green line. This reveals that the Pareto optimal solution is given by
two different mechanisms: in the left part of
Fig.~\ref{fig:Pareto_fronts}b, the Pareto-optimal solution is
determined by the more efficient driving, whereas in the
right part it is dominated by the lowest dissipation. When both
curves coincide, the Pareto front transitions smoothly between the two
limiting cases. 

We conclude that the shape of the Pareto front is dominated by
the solution sets $a(\gamma) = (k-\gamma)^2/4\gamma$ and $a =\gamma
\sigma_x^2/2k\langle y^2\rangle_{\rm 0}$. These lines characterize the adaptation
ability of the feedback cycle and the distance from the initial state,
respectively. In Appendix~\ref{app:additional_plots} we provide more
data to support the conclusion. There, we consider different initial
values and discuss the role of the initial state $\langle
y_0^2\rangle$.

\subsection{Adaptation to harmonic perturbations}
\label{subsec:feedback_2}

Molecular process are often subject to periodic
perturbations
from the environment, for example, by daily light or temperature variations
\cite{Mehra2006Jul, dobroborsky2006thermodynamics}. The biological
response to such periodic external modulations is known as \textit{circadian
  rhythms}. There exists diverse physiological mechanisms, known as
biological clocks, that establish a time-periodic response to external
perturbations. Heat-shock proteins are somewhat different, they do \emph{not} exhibit self-sustained oscillations but are activated by slow
time-periodic environmental changes. The daily periodic activation of
HSP90 has been found to regulate the behavior of actual biological
clocks, which are sensitive to temperature \cite{Goda2014Oct}. We
focus on the response of the model thermostat to such a time-periodic temperature
variation in the long-time limit. We will show that for times larger
than the relaxation time, the system reaches a periodic state where
the molecule $y$ is continuously activated by the distortion. 

We consider that $x$ is subjected to a harmonic temperature variation
with frequency $\Omega$, which may represent, e.g.\ daily temperature oscillations,
\begin{equation}
    \sigma_x^2(t) = \sigma_{x0}^2 + \delta \cos(\Omega t)
\end{equation}
where the amplitude obeys $\delta < \sigma_{x0}^2$. Since we are only
varying the temperature, the distribution remains a zero-mean Gaussian with
a covariance evolving according to
\begin{equation}
    \dot{\textbf{v}}_t = M\textbf{v}_t + \bchi(t), \label{eq:lyapunov_on_time}
\end{equation}
where now $\bchi(t) = \bchi_0 + \bchi_{\rm p}(t)$ is the a sum of a
constant part $\bchi_0 = (\sigma_{x0}^2, 0, \sigma_{x0}^2/ka)$ and a
harmonic part $ \bchi_{\rm p}(t) = (\delta \cos(\Omega t), 0, \delta
\cos(\Omega t)/ka)$. Eq.~\eqref{eq:lyapunov_on_time} has the solution
\begin{equation}
    \textbf{v}_t = \ee{Mt}\textbf{v}_0 + \ee{Mt}\int_0^t \ee{-Mt'}\bchi(t') dt'.
    \label{eq:preliminar_solution}
\end{equation}
We can in turn choose $\sigma_y^2(t)$ in several ways, depending on how we
want to interpret the adaptation in the time-dependent context. The
arguably most intuitive (and consistent) choice seems to be to simply impose
$\sigma_y^2(t) = \sigma_x^2(t)/k a$ at any time, i.e.\ $y$ up to the
amplitude $y$ experiences the same temperature as $x$. 
Then, the instantaneous quasi-steady state of Eq.~\eqref{eq:lyapunov_on_time}, given by $0=\dot{\textbf{v}}_t = M\textbf{v}^{\rm qs}_t + \bchi(t)$ at any $t$, satisfies $\langle x_t^2\rangle_{\rm qs} = \sigma_x^2(t)/2k$ and $ \langle y_t^2\rangle_{\rm qs} = \gamma\sigma_x^2(t)/2ka$.

We can use the property $\ddot{\bchi}_{\rm p}(t) = -\Omega^2 \bchi_{\rm p}(t)$ to
integrate Eq.~\eqref{eq:preliminar_solution} by parts, yielding
\begin{align}
        \textbf{v}_t &= \ee{Mt}\textbf{v}_0 - \left(1-\ee{Mt}\right)M^{-1}\bchi_0
        \\
        &+\left(1+\Omega^2M^{-2}\right)^{-1}\left[-M^{-1}\bchi_{\rm p}(t) - M^{-2} \dot{\bchi}_{\rm p}(t) \right]\nonumber
        \\
        & +
        \left(1+\Omega^2M^{-2}\right)^{-1}\ee{Mt}\left[M^{-1}\bchi_{\rm p}(0) +
          M^{-2} \dot{\bchi}_{\rm p}(0)\right].\nonumber
        \label{aux_chi}
\end{align}
Finally, in the long time regime $\ee{Mt} \overset{t\to\infty}{=} 0$ where
correlations with the initial state of the system vanish, we obtain
\begin{equation}
    \begin{split}
        \textbf{v}_t & \overset{t\to\infty}{=} -M^{-1}\bchi_0
         \\
        &-\left(1+\Omega^2M^{-2}\right)^{-1}\left[M^{-1}\bchi_{\rm p}(t) + M^{-2} \dot{\bchi}_{\rm p}(t) \right],
    \end{split}\label{eq:long_time_dynamics}
\end{equation}
where the covariance vector $\textbf{v}_t$ exhibits an $\Omega$-dependent
oscillatory behavior.

\begin{figure*}
    \centering
    \includegraphics[width=1\linewidth]{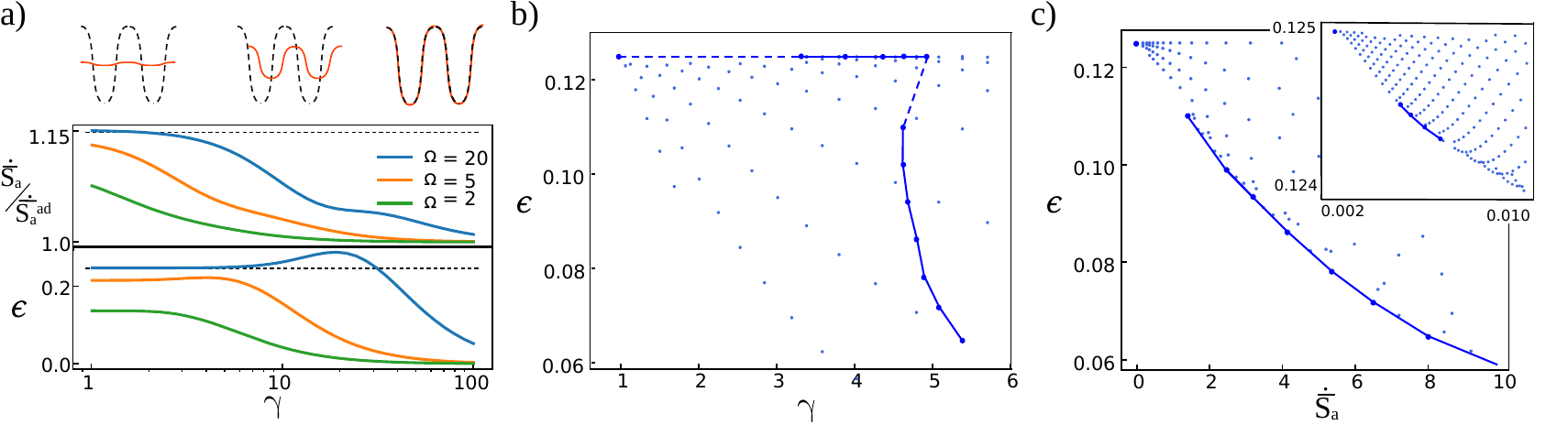}
    \caption{(a)~Entropy production per cycle $\dot{\Bar{S}}_{\rm a}$
      relative to the adiabatic limit $\dot{\Bar{S}}_{\rm a}^{\rm ad}$
      and adaptation error $\epsilon(\tau)$ as a function of $\gamma$
      and for several values of $\Omega$.~Black dashed lines represent
      the high-$\Omega$ limit with $\dot{\Bar{S}}_{\rm a}^{\rm
        ad}/\dot{\Bar{S}}_{\rm a} =1/ \sqrt{1 -
        \delta^2/\sigma_{x0}^4} \approx 1.15$ and $\epsilon(\tau) =
      \delta/2k = 0.25$. We set $a = (k-\gamma)^2/4\gamma$.~Top:~Schematic of different classes of responses to
      harmonic distortions; the black line represents the
      instantaneous value of $\sigma_x^2(t)$ and the red line the
      response $\langle x_t^2\rangle$.~(b)~Adaptation error $\epsilon$
      as a function of $\gamma$ for several values of $a\in [10^{-3},
        2]$ and $\gamma\in [1, 6]$, with $\Omega = 2$.~The continuous
      line depicts the Pareto front,~dashed lines represent
      discontinuities of the Pareto front due to concave regions and are merely a guide
      for the eye.~(c) Adaptation error $\epsilon$ as a function of
      $\dot{\Bar{S}}_{\rm a}$ for the same values of $a,
      \gamma,\Omega$ as in (b).~Inset:~Magnification of the small-$\dot{\Bar{S}}_{\rm a}$ region.~The results correspond
      to $k = \sigma_{x0}^2 = 1$, $\delta = 0.5$.}
    \label{fig:harmonic}
\end{figure*}

For high perturbation frequencies, $\Omega\gg 1$, the second term is
$\mathcal{O}(\Omega^{-2})\ll 1$ and the system effectively reaches a stationary
state given by $\textbf{v}_{\rm s}\simeq M^{-1}\bchi_0$ with $\langle
x^2 \rangle_{\rm s} = \sigma_{x0}^2/2k$ and $\langle y^2 \rangle_{\rm s} = \gamma\sigma_{x0}^2/2ka$.
This situation is shown schematically in Fig.~\ref{fig:harmonic}a (top left).

In the opposite limit, which we refer to as the \textit{adiabatic limit}, the system is able to perfectly adapt to
the environmental perturbations at any time (see
Fig.~\ref{fig:harmonic}a, top right). Precisely, we impose
$M^{-2}\to 0$ as well as $\Omega^2 M^{-2}\to 0$, that is, that the
relaxation is fast and in particular much faster than the
oscillation period $\Omega^{-1}$. In this adiabatic regime 
we find $\textbf{v}^{\rm ad}_t =
-M^{-1}[\bchi_0 +\bchi_{\rm p}(t)]$.

Meanwhile, in the intermediate
regime (see Fig.~\ref{fig:harmonic}a, top center) the adaptation is
imperfect and $\langle x_t^2 \rangle$ exhibits a time-periodic
evolution with varying amplitude, delayed with respect to the
environmental perturbation.

We define the adaptation error analogously to
Eq.~\eqref{eq:adaptation_error_definition} and rewrite it using Eq.~\eqref{eq:long_time_dynamics} as
\begin{equation}
\begin{split}
    \epsilon(t) =& |\langle x_t^2\rangle - \sigma_x^2(t)/2k|
    \\
    =& \left|\textbf{e}_0^T
    \left(1+\Omega^2M^{-2}\right)^{-1}
    M^{-2}\dot{\bchi_{\rm p}}(t) \right.\\
    +&\,\textbf{e}_0^T\left.\left[
    \left(1+\Omega^2M^{-2}\right)^{-1} - 1
    \right]
    M^{-1}\bchi_{\rm p}(t)
    \right|.
\end{split}
\label{adapted}
\end{equation}
As before, the error $\epsilon(t)$ indicates how far the controlled
system $x_t$ is at time $t$ from the perfectly adapted instantaneous
situation. In the adiabatic limit, the adaptation is perfect for any
$t$, so $\epsilon(t) = 0$. In the limit of high $\Omega$
Eq.~\eqref{adapted} instead reduces to
\begin{equation}
    \epsilon(t) \overset{\Omega \gg \gamma}{\simeq} |\textbf{e}_0^TM^{-1}\bchi_{\rm p}(t)| = \delta/2k|\cos(\Omega t)|.
\end{equation}
In the large-time limit the instantaneous adiabatic entropy production in
Eq.~\eqref{eq:adiabatic_entropy}  is also a time-periodic function. The
adiabatic entropy produced \emph{per period} is
\begin{equation}
    \dot{\Bar{S}}_{\rm a} \equiv \frac{1}{\tau}\int_0^\tau
    \dot{S}_{\rm a}(t) dt,
    \label{period}
\end{equation}
where $\tau = 2\pi/\Omega$ is the period of the
distortion. Using Eq.~\eqref{eq:adiabatic_entropy} in the adiabatic regime we find
\begin{equation}
    \dot{\Bar{S}}_{\rm a}^{\rm ad} = a\left(1+\frac{\gamma}{k}\right).\label{eq:adiabatic_prediction}
\end{equation}
Evaluating instead the integral in Eq.~\eqref{period} in the high-$\Omega$
limit we find
\begin{equation}
     \dot{\Bar{S}}_{\rm a} \overset{\Omega \gg \gamma}{\simeq}  \frac{
       \dot{\Bar{S}}_{\rm a}^{\rm ad}}{\tau}\int_0^\tau
     \frac{\sigma_{x0}^2}{\sigma_{x}^2(t)} dt =
     \frac{\dot{\Bar{S}}_{\rm a}^{\rm ad}}{\sqrt{1 -
         \delta^2/\sigma_{x0}^4}}\ge   \dot{\Bar{S}}_{\rm a}^{\rm ad}.\label{eq:high_Omega_prediction}
\end{equation}
The adiabatic entropy production $\dot{\Bar{S}}_{\rm a}$ and
adaptation error $\epsilon(\tau = 2\pi/\Omega)$ for several values of
$\gamma$ and $\Omega$ are shown in Fig.~\ref{fig:harmonic}a. We observe for $\Omega = 20$ and $\gamma \simeq 1$ that both,
$\dot{\Bar{S}}_{\rm a}$ and $\epsilon(\tau)$ remain close to the
high-frequency limit in Eq.~\eqref{eq:high_Omega_prediction} with
$\epsilon(\tau) = \delta/2k$ (dashed line in
Fig.~\ref{fig:harmonic}a). The high-frequency limit is, however,
\emph{not} reached for smaller values $\Omega = 5 ,2$, since these are
comparable to the minimum $\gamma = 1$. Increasing $\gamma$ we
gradually move into the adiabatic regime in
Eq.~\eqref{eq:adiabatic_prediction} with $\epsilon(t) = 0$ for any
$\Omega$. 
Interestingly, the dependence of $\epsilon(\tau)$ on $\gamma$ is
non-monotonic;~it exhibits a maximum between the high-$\Omega$ and the adiabatic limits.

In Fig.~\ref{fig:harmonic}b,c we show $\epsilon(\tau)$ for several
values of $a$ and $\gamma$ whereby $\Omega = 2$. These results are
similar to those in Fig.~\ref{fig:Pareto_fronts}, in particular
Fig.~\ref{fig:harmonic}c reveals a clear border between reachable and
impossible values of $\epsilon$ and $\dot{\Bar{S}}_{\rm a}$, i.e.\ the
Pareto front. Unlike the example in Fig.~\ref{fig:Pareto_fronts}, the Pareto front here
exhibits a concave region at low $\dot{\Bar{S}}_{\rm a}$.

The convex part of the front can be reconstructed using the method in
Eqs.~(\ref{conv}-\ref{inf}), i.e. finding the values of $a$ and
$\gamma$ that minimize $J(a,\gamma; \alpha) = \alpha \epsilon(a,
\gamma, \tau) + (1-\alpha)\dot{\Bar{S}}_{\rm a}(a, \gamma, \tau)$ for
$\alpha \in (0,1)$. The results of this method are shown in
Fig.~\ref{fig:harmonic}b,c as points connected by the continuous blue
lines. As shown in Fig.~\ref{fig:harmonic}b, there are in fact three
disjoint convex sections: a point (top left), the horizontal line, and the
vertical curved line. While the former two are well separated in the
$(\epsilon,\gamma)$ space in Fig.\ref{fig:harmonic}b,  they cannot be
distinguished in Fig.~\ref{fig:harmonic}c.

The disjoint sections of the Pareto front reveal distinct underlying
optimization mechanisms \cite{Seoane2016May}. The isolated point at
$\gamma = 1,\epsilon(\tau) = 0.125$ represents the limit in which $y$
has \emph{no} effect on $x$, dissipating no energy. Along the
horizontal segment in Fig.~\ref{fig:harmonic}b the Pareto-optimal
solution increases $\gamma$ without improving $\epsilon$
significantly, since the system is stuck in the high-$\Omega$
regime. The vertical segment in Fig.~\ref{fig:harmonic}b represents
the optimization close to the adiabatic regime, where both $\gamma$ and
$a$ can be increased to approach the adiabatic regime. The jump
between both segments is a result of the non-monotonic dependence of
$\epsilon$ on $\gamma$.

\subsection{Stochastic temperature shocks}
\label{subsec:feedback_3} 

In certain situations biological processes are continuously stimulated
by successive, abrupt changes in environmental conditions at discrete
instances of time \cite{Lan2012May}. Each stimulus triggers the
adaptation mechanism during a relaxation period. We now consider the
evolution of the concentration $x_t$ in such situations.

To represent the random stimulus, we consider again $x_t$
following Eq.~\eqref{eq:langevin} with time-independent
$\sigma_x$. At random times $t_i$, both $x_t$ and $y_t$ are
instantaneously reset to predefined values distributed according to
the probability density $p(x,y)_{\rm res}$. The joint probability
density of $x_t$ and $y_t$ is hence reset at all $t_i$ according to
$p(x,y;t_i) \rightarrow p(x,y)_{\rm res}$ and henceforth evolves from
$p(x,y)_{\rm res}$. Times $t_i$ are ssumed to be Poisson
distributed with rate $\Gamma$. 

The joint probability density of $x_t,y_t$, $p_t(\textbf{x})\equiv p_t(x,y)$, follows the extended Fokker-Planck equation \cite{Mori2023May}
\begin{equation}
    \partial_t p_t(\textbf{x}) = -\nabla\cdot \textbf{J}_t(\textbf{x}) + \Gamma\left[p_{\rm res}(\textbf{x})-p_t(\textbf{x})\right], \label{eq:reseting}
\end{equation}
with $\textbf{J}_t(\textbf{x}) = -A\textbf{x}p_t(\textbf{x}) - D\nabla
p_t(\textbf{x})$ and $A$ given in Eq.~\eqref{AA} and $D\equiv{\rm  diag}\,( \sigma_x^2/2,\gamma^2\sigma_y^2/2)$. 
The solution $p_t(\textbf{x})$  is not Gaussian, not even if we choose
a Gaussian resetting density $p_{\rm res}(\textbf{x})$. However, the
vector of
covariance elements evolves according to
\begin{equation}
    \dot{\textbf{v}}_t = M\textbf{v}_t + \textbf{d} + \Gamma
        [\textbf{v}_{\rm res} - \textbf{v}_t],
        \label{cov_res}
\end{equation}
where $\textbf{d} \equiv (\sigma_x^2, 0,
\gamma^2\sigma_y^2)^T$, the matrix $M$ was defined in
Eq.~\eqref{MM}, and $\textbf{v}_{\rm res} \equiv (\langle
x^2\rangle_{\rm res}, 0, \langle y^2\rangle_{\rm res})$ is the
covariance of $p_{\rm res}(\textbf{x})$ written in vector form for
notational convenience. 

In contrast to the previous examples, where we focused on the
adaptation during the transient (time-dependent regimes), we here observe how the
system reaches a non-equilibrium steady state satisfying
\begin{align}
\textbf{v}_{\rm NESS} = -(M - \Gamma\mathbb{I})^{-1}(\textbf{d} +
\Gamma \textbf{v}_{\rm res}),
\label{v_ness}
\end{align}
with $\mathbb{I}$ being the $3 \times 3$ identity matrix. In
Appendix~\ref{app:perturbation} we derive two  asymptotic
solutions of Eq.~\eqref{v_ness}. For low rates $\Gamma\ll 1$ we have
$\textbf{v}^-_{\rm NESS} \simeq \textbf{v}_{\rm s} + \Gamma
M^{-1}(\textbf{v}_{\rm s} - \textbf{v}_{\rm res})$, i.e.\ the system
remains close to the perfectly adapted state $\textbf{v}_{\rm
  s}$. Conversely, for large resetting rates $\Gamma\gg 1$, we find
$\textbf{v}^+_{\rm NESS} \simeq \textbf{v}_{\rm res} +
M(\textbf{v}_{\rm res}-\textbf{v}_{\rm s})/\Gamma$, that is,
$\textbf{v}_t$ is pushed to the reset state $\textbf{v}_{\rm res}$. 

We define the adaptation error as
\begin{align}
    \epsilon(a, \gamma) &\equiv |\langle x^2\rangle_{\rm NESS} - \langle x^2\rangle_{\rm s}| 
    \\
    &= |\textbf{e}_0^T\left[(M - \Gamma\mathbb{I})^{-1}(\textbf{d} + \Gamma \textbf{v}_{\rm res}) + M^{-1}\textbf{d}\right]|, \nonumber
\end{align}
which is time independent. In the two asymptotic regimes we obtain, respectively,
\begin{align}
    \epsilon(a, \gamma) &\overset{\Gamma \gg k}{\simeq}
    \left|\textbf{e}_0^T  \left(\mathbb{I} - M/\Gamma \right)(\textbf{v}_{\rm res}-\textbf{v}_{\rm s}) \right|\\
    \epsilon(a, \gamma) &\overset{\Gamma \ll k}{\simeq} \Gamma \left|\textbf{e}_0^T M^{-1}(\textbf{v}_{\rm s} - \textbf{v}_{\rm res}) \right|.
    \label{eq:high_Gamma}
\end{align}
In the non-equilibrium steady state the total entropy production is
time independent and is comprised of $\dot{S}_{\rm a}$ due to the
temperature difference between the subsystems and a contribution of
the  resetting, proportional to $\Gamma$ \cite{Mori2023May}. 
 However, for a fixed $\Gamma$, only the adiabatic entropy production $\dot{S}_{\rm a}$ plays a relevant
 role in the trade-off between adaptation and dissipation, such that
 we may, without loss of generality, 
 consider this contribution only.

 As already mentioned, $p_t(\textbf{x})$ in Eq.~\eqref{eq:reseting}
 is not Gaussian and Eq.~\eqref{eq:adiabatic_entropy} hence does \emph{not}
 apply. 
 We evaluate $\dot{S}_{\rm a}$ numerically for a Gaussian $p_{\rm
   res}(\textbf{x})$, exploiting that between any
two consecutive resets at times $t_i$ and $t_{i+1}$, the process is
indeed Gaussian. The initial condition for each trajectory is drawn
from the NESS of the non-reset process. 
That is, between any such $t_i$ and $t_{i+1}$ the
adiabatic entropy production follows
Eq.~\eqref{eq:adiabatic_entropy} with $\textbf{v}_{t_i}=\textbf{v}_{\rm res}$. We denote such contributions for a
given sample sequence of $\{t_i\}(\omega)$ by $\dot{S}_{\rm a}(t|i)$, leading to
\begin{align}
\Delta S^t_{\rm a}(\omega)\equiv \sum_{i=0}^{N(\omega)}\int_{t_i}^{t_{i+1}} \dot S_{\rm a}(t|i) dt,
\label{snippet}
\end{align}
where $N(\omega)$ is the number of resets in the sample sequence
$\omega$ and we set $t_{N(\omega)+1}\equiv t$. 
Note that these account
for the average over Langevin evolutions~\eqref{eq:langevin} between
the resets but \emph{not} yet over different realizations of reset
times. We now simulate $10^3$ statistically independent trajectories
of reset times of length $t=120$ setting $\langle x^2\rangle_{\rm res}
 = \sigma_x^2/2k$, and $\langle y^2\rangle_{\rm res} =
 \gamma\sigma_y^2/2$. 
The adiabatic entropy production in $[0,t]$ then follows as
an average over all samples $\Delta S^t_{\rm a}=\langle \Delta
S^t_{\rm a}(\omega)\rangle_\omega$. 

\begin{figure*}
    \centering
    \includegraphics[width = 1\linewidth]{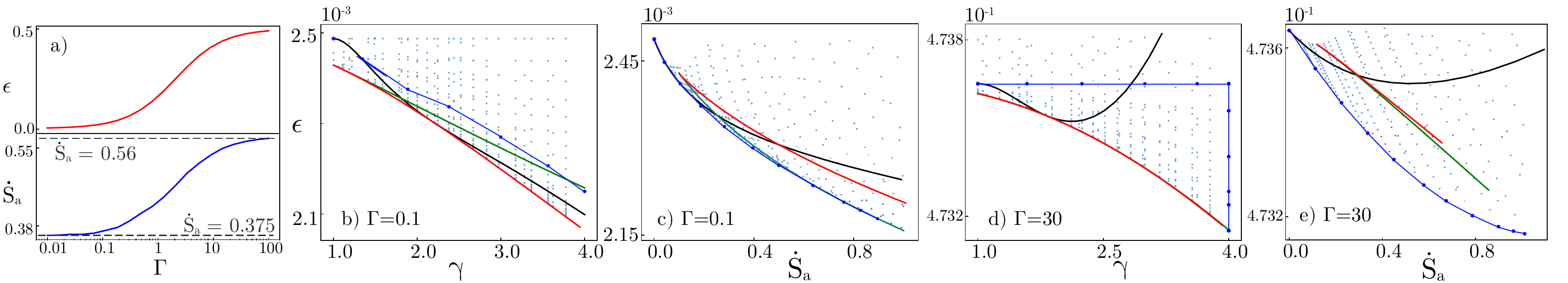}
    \caption{(a)~Adaptation error and adiabatic entropy production
      on $[0,t]$ as a function of $\Gamma$ for $\gamma = 2$ and $a =
      (k-\gamma)^2/4\gamma$.~(b-e)~Adaptation error as a function of
      $\gamma$ and $\dot{S}^t_a$ for $\Gamma = 0.1$ (b-c) and $30$
      (d-e). Each point represent a different value of $a \in
      [10^{-3}, 2]$ and $\gamma \in [1,4]$. Red lines denote the
      minimal adaptation error, the blue curves are the Pareto fronts,
      and black and green curves are defined by $a =
      (k-\gamma)^2/4\gamma$ and $a = \gamma\sigma_x^2/2k\langle
      y^2\rangle_0$, respectively.~The remaining parameters are the same as in Fig.~\ref{fig:Pareto_fronts}.}
    \label{fig:Adaptation_figure}
\end{figure*}

The results for $\epsilon$ and $\dot{S}^t_{\rm a}$ obtained this way are shown in Fig~\ref{fig:Adaptation_figure}a for different values of
$\Gamma$. According to the prediction~\eqref{eq:high_Gamma}, the
adaptation error vanishes for low $\Gamma$ and increases monotonically
up to  $\epsilon = \langle x^2\rangle_{\rm res} - \sigma_x^2/2k = 1/2$
for $\Gamma \simeq 100$. Similarly, $\dot{S}^t_{\rm a}$ a increases
monotonically with $\Gamma$. Interestingly, for low $\Gamma$ we find
$\dot{S}^t_{\rm a} = 0.357$, whereas for high $\Gamma$ we recover
$\dot{S}^t_{\rm a} = 0.56$, corresponding to the Gaussian prediction in
Eq.~\eqref{eq:adiabatic_entropy} with $\textbf{v} = \textbf{v}_{\rm
  s}$ and $\textbf{v} = \textbf{v}_{\rm res}$, for the respective limits,
suggesting that the limiting distribution is essentially Gaussian.

In Fig.~\ref{fig:Adaptation_figure}b-e we analyze $\epsilon$ and
$\dot{S}^t_{\rm a}$ for several values of $\gamma$ and $a$ in the
intermediate regimes $\Gamma = 0.1, 30$.  In the first case (see Fig.~\ref{fig:Adaptation_figure}b,c)
the behavior  is similar to that in
Fig.~\ref{fig:Pareto_fronts}. Indeed, for low $\Gamma$ the system has
enough time to equilibrate upon a reset, and we can explain the Pareto
front (blue line) 
and the lowest error curve (red line) using the curves $a =
(k-\gamma)^2/4\gamma$ (black line) and $a = \gamma\sigma_x^2 /
2k\langle y^2\rangle_{\rm res}$ (green line), as in the
beginning of the section.  

The situation changes for $\Gamma = 30$ (see
Fig.~\ref{fig:Adaptation_figure}d-e), since $x_t$ has \emph{no} time
to recover between resetting events and the feedback action of $y_t$
is \emph{not} relevant. In Fig.~\ref{fig:Adaptation_figure}d we observe
that the minimum error is obtained when $y_t$ is reset into its own
equilibrium state (i.e.\ the green and red lines in
Fig.~\ref{fig:Adaptation_figure}d coincide).
The Pareto front is again determined by defining $J(a, \gamma; \alpha)
\equiv \alpha\epsilon(a, \gamma) + (1-\alpha)\dot{S}^t_{\rm a}(a,
\gamma)$ and finding the minima for $\alpha \in (0,1)$. In
Fig.~\ref{fig:Adaptation_figure}e we observe that the front is convex
in the entire parameter space, and coincides at $\gamma = 1$ with the
optimal entropy production (black curve in Fig.~\ref{fig:Adaptation_figure}e).
In  Fig.~\ref{fig:Adaptation_figure}d we find that the Pareto front becomes vertical at $\gamma =
4$, indicating that the lower part of the front in
Fig.~\ref{fig:Adaptation_figure}e is bounded by the limiting maximum value $\gamma = 4$. In this case, the Pareto front depends on our choice of the parameter space, so one could continue improving the solutions by increasing $\gamma$.

\section{Receptor-ligand reaction network}\label{sec:chemical}

\textcolor{black}
{Certain fundamental biological processes, such as biosynthesis and
biomolecular recognition \cite{Qian2007May}, are underpinned by
receptor-ligand reaction networks. In protein biosynthesis such networks, sketched in
Fig.~\ref{fig:chemical_dynamics}a, represent the binding of an amino
acid (or more generally \textit{ligand}) to an enzyme-substrate
complex (referred to here simply as the \textit{receptor}) to form a
peptide bond and thus expand the synthesized protein (which with a
slight abuse of terminology we will call
\textit{complex}). The synthesis may be coupled to an ATP-hydrolysis
reaction that provides a free energy input to drive the ligand binding
out-of-equilibrium, thereby promoting the formation of a ``correct'' complex
\cite{Boeger2022Jun, Qian2007May}. 
A celebrated application is known is kinetic proofreading, whereby the
free energy released by ATP hydrolysis is utilized to discriminate the
incorporation of ``correct'' ligands in a mixture of ligands
\cite{Boeger2022Jun}. Namely, intact protein synthesis requires a precise recognition of correct amino acids by the ribosome. However, many amino acid are similar, giving rise to synthesis errors. More generally, substrates can react with different undesired ligand molecules at the same time, creating spurious complex products.
Furthermore, receptor-ligand networks can regulate a cell's response
to different types of chemical stimuli, by acting as very sensitive
switches triggered by concentration changes \cite{Qian2005Jan,
  Goldbeter1981Nov}. Such switches  play a fundamental role in adaptation \cite{Koshland1982Jul}.}

\color{black}Here we study, in a generic minimal context, the adaptation
of a receptor-ligand network upon a change of concentration of the ligand
converted to a fixed amount of the complex. We determine
optimal solutions to the adaptation-dissipation trade-off, revealing
distinct mechanisms underlying optimal solutions.
We consider the reversible
reaction between a receptor $R$ and the ligand $L$ to create a complex
$R_L$. The activation of this complex is coupled to the hydrolysis
of ATP (denoted $T$) to ADP (denoted $D$), $R_L +
T\rightarrow R_L^* + D$. In this section, we will consider the 
transient initiated when ligand $L$ arrives to $R$. When this
happens, the network relaxes into an adapted state defined by the
concentration of $R_L$. The velocity of the relaxation into the
adapted state is given by the reaction rates $k_{1,2,3}$ and the
concentrations of $T, D$ and $L$. With a slight abuse (albeit
simplification) of notation we use  $T, D$ and $L$ to denote the
concentrations of the respective molecules.

\color{black}

We assume the environment to be at constant temperature, and moreover
consider chemostated (i.e.\ constant) concentrations of $L, D$ and
$T$. The evolution of the probability (or concentration) of receptor and both complex states
follows the master equation $\dot{\textbf{p}}_t=M_{\rm R}\textbf{p}_t$
with \cite{Qian2007May}
\begin{equation}
   M_{\rm R}\equiv\left(
    \begin{matrix}
        - L k^+_1 - Lk^-_3 & k^-_1 & k^+_3\\
        Lk^+_1 & - Tk^+_2 - k^-_1 & Dk^-_2\\
        L k^-_3  & T k^+_2 &  -Dk^-_2 - k^+_3
    \end{matrix}
    \right)\label{eq:chemical_dynamics}
\end{equation}
where $\textbf{p}_t = [p_R(t), p_{R_L}(t), p_{R_L^*}(t)]^T$ is the
probability of the respective states at time $t$, normalized according
to $p_R(t) + p_{R_L}(t) + p_{R_L^*} = 1$. The species inter-convert with
chemical rates $k^+_i$, $i = 1,2,3$ as denoted by the black arrows in
Fig.~\ref{fig:chemical_dynamics}a.
The reverse reactions occur with rates $k^-_i = k^+_i\exp(\beta\Delta \mu_i)$, where $1/\beta$ is the environmental temperature. $\Delta \mu_i$ is the chemical potential difference of each reaction, satisfying $\Delta \mu_1 + \Delta \mu_2 + \Delta \mu_3 = 0$. The total chemical potential driving the reaction is
\begin{equation}
    \beta\Delta \mu = \ln\left(T/D\right).
\end{equation}
In contrast to the feedback system in the previous section, which was
driven by a temperature difference,  the thermodynamic driving force
here is the concentration difference between $T$ and $D$. If $T = D$
the system evolves towards thermal equilibrium with $p_R^{\rm eq},
p_{R_L}^{\rm eq}, p_{R_L^*}^{\rm eq}$ satisfying the detailed balance
conditions, $p_{R_L}^{\rm eq}/p_{R}^{\rm eq} = \exp(-\beta \Delta
\mu_1), p_{R_L^*}^{\rm eq}/p_{R_L}^{\rm eq}= \exp(-\beta \Delta
\mu_2)$ and $p_{R}^{\rm eq}/p_{R_L^*}^{\rm eq} = \exp(-\beta \Delta
\mu_3)$, independently of $L$.  If, however, $\Delta \mu \neq 0$, the
hydrolysis reaction will create steady-state currents in the reaction
cycle, with a sign dictated by the relative concentrations of $T$ and
$D$.  These currents produce entropy proportional to $\Delta \mu$ \cite{Qian2007May}.

\subsection{Adaptation}

The stationary probability distribution of the network  is set by the
concentration of $R_L$, and must be independent of other parameters of
the dynamics.  We choose the adaptation condition
\begin{equation}
   p_{R_L}^{\rm st}/p_{R}^{\rm st} = \phi = \rm{const.}
    \label{eq:chemical_constraint}
\end{equation}
where $\phi$ is a fixed parameter.
Imposing  $\dot{\textbf{p}}_t = 0$ and inserting
Eq.~\eqref{eq:chemical_constraint} into the master equation we obtain the following relation between $L$, $T$ and $D$
\begin{equation}
    \frac{L}{\phi} = \frac{k^-_1k^+_3 + Dk^-_1k^-_2 + Tk_3^+k^+_2
    }{k^+_1k^+_3 + Dk^+_1k^-_2 + Dk^-_3k^-_2 },
    \label{eq:L_constraint}
\end{equation}
i.e.\, the constant concentrations of $T$ and $D$ determine the
concentration of $L$ that is required to satisfy the
condition in Eq.~\eqref{eq:chemical_constraint}.

\begin{figure*}[htb!]
    \centering\includegraphics[scale=0.5]{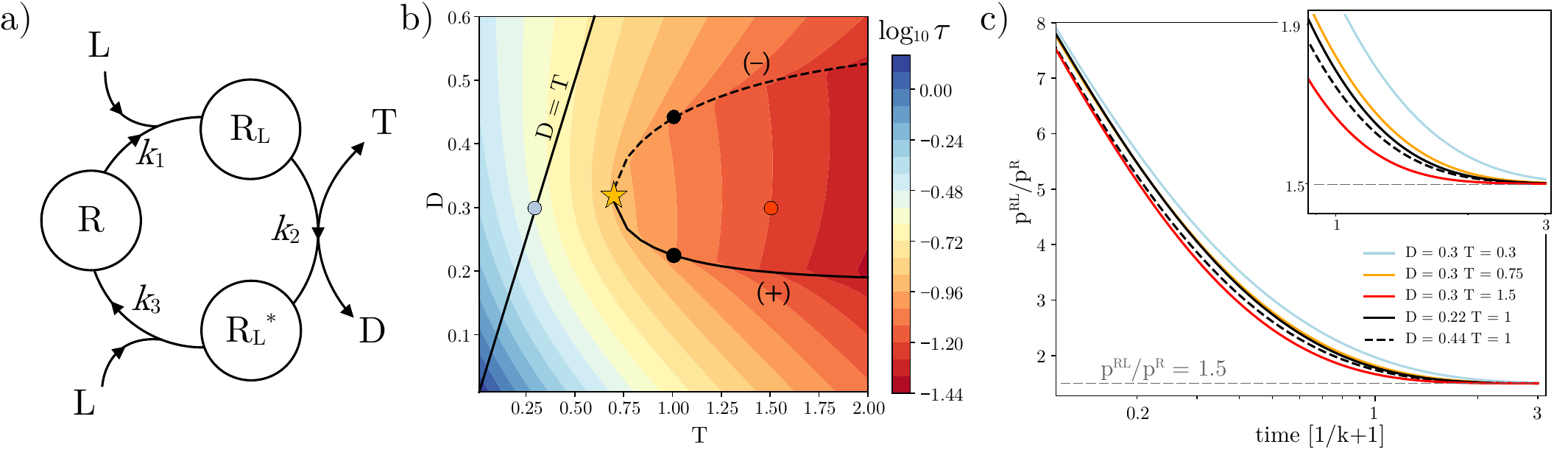}
    \caption{(a)~\color{black}Schematic of receptor-ligand chemical reaction network.\color{black}~(b) $\log_{10} \tau$ (color bar) as a function of $T$ and
      $D$. The black lines are given by $D_+(T)$, $D_-(T)$ and
      $D=T$. The blue and red circles and the star represent the
      points used considered in panel c.~(c)~Probability ratio~$P_{R_L}(t)/P_{R}(t)$
      for several values of $T$ and $D$. The inset shows a
      magnification of the long-time regime.~We used $\phi = 1.5$,
      $\beta = 1$, $k^+_1 = k^+_2 = k^+_3 = 1$, $\Delta \mu_1 = \Delta
      \mu_3 = -0.5$ and $\Delta \mu_1 = 1$. $L$ is given by
      Eq.~\eqref{eq:L_constraint}. $p_{R}(0) = 0.1$, $p_{R_L}(0) = 0.9$
      and $p_{R_L}(0) = 0$.} 
    \label{fig:chemical_dynamics}
\end{figure*}

The solution of the master equation is $\textbf{p}_t = \ee{M_
R t}\textbf{p}_0$, where $M_R$ is defined in
Eq.~\eqref{fig:chemical_dynamics}.
The matrix $M_{\rm R}$ has one zero eigenvalue, whose right eigenvector is
the steady-state probability already described by
Eq~\eqref{eq:chemical_constraint}. The remaining two eigenvalues,
$\lambda_+$ and $\lambda_-$ describe the evolution during the
transient. Their real parts  $\rm{Re}[\lambda_\pm] < 0$ are negative, whereas the imaginary parts
$\rm{Im}[\lambda_\pm]$ may (but does not need to) be different from
zero. If it is non-zero, it gives rise to oscillations during the
transient.

The relaxation timescale of the
transient is given by $\tau =
1/\min(\rm{Re}[\lambda_+],\rm{Re}[\lambda_-])$ and is shown in  
Fig.~\ref{fig:chemical_dynamics}b (on a logarithmic scale) for
several values of $T$ and $D$. It depends explicitly on $\phi$ due
to Eq.~\eqref{eq:chemical_constraint}. The full and dashed black  lines denoted by $(\pm)$
represent concentrations for which $\rm{Im}[\lambda_\pm]$
becomes different from zero, and enclose the region of oscillatory
transients.  From
Fig.~\ref{fig:chemical_dynamics}b we conclude that, increasing $T$ above $D$ accelerates
adaptation by shortening the transient. This gives rise to a clockwise
steady-state circulation in
the reaction cycle in Fig.~\ref{fig:chemical_dynamics}a.
Counter-clockwise oscillations
imply moving across the line $D = T$ and only marginally improve $\tau$. For
low values of $T$ and $D$, the timescale $\tau$ decays along the $D =
T$ line due to the increasing concentration of $L$.

The system's adaptation speed to $\phi = 1.5$ is shown in Fig.~\ref{fig:chemical_dynamics}c
for several values of $T$ and $D$ denoted in
Fig.~\ref{fig:chemical_dynamics}b with red/blue points and a
star. The initial condition is $p_{R}(0) = 0.1$, $p_{R_L}(0) = 0.9$ and $p_{R_L}(0) = 0$. Notably, we observe \emph{no} discernible
oscillations for parameters chosen within the nominally oscillatory
region where $\rm{Im}[\lambda_\pm]\ne 0$.

\subsection{Adaptation-dissipation trade-off}\label{subsec:chemical_1}

As before we characterize the adaptation accuracy in terms of the
\textit{adaptation error} here defined as
\begin{equation}
  \epsilon (T, D; t) = |p_{R_L}(t)/p_{R}(t) - \phi|
  \label{def_error_n}
\end{equation}
at any time $t$. For times much larger than the relaxation time
$t\gg\tau$ the system adapts to the desired steady state such that
$\epsilon\to 0$. In the following, we consider the error at $t = 1/k^+_{1,2,3} =
1$, when the system is not yet completely relaxed as seen in
Fig.~\ref{fig:chemical_dynamics}c. 

To describe dissipation, we consider the adiabatic entropy production rate, given by \cite{Esposito2010Jul}
\begin{equation}
    \dot S_{\rm a} (T, D; t) = \sum^3_{i,j=1} M_{R}^{ij}(\mathbf{p}_t)_j\ln\left(\frac{ M_{R}^{ij}(\mathbf{p}^{\rm st}
    )_j } {M_{R}^{ji} (\mathbf{p}^{\rm st})_i }\right) \geq 0.
    \label{eq:chemical_ent_production}
\end{equation}
where $(\textbf{p}_t)_i$ for $i=1,2,3$ are the components of
$\textbf{p}_t$ and $\mathbf{p}^{\rm st}=\textbf{p}_{t\to\infty}$.~The entropy production rate is positive during the
transient as well as in the non-equilibrium steady state. It only
vanishes at equilibrium, attained for $T = D$ for which detailed balance holds.

\begin{figure*}
    \includegraphics[scale = .6]{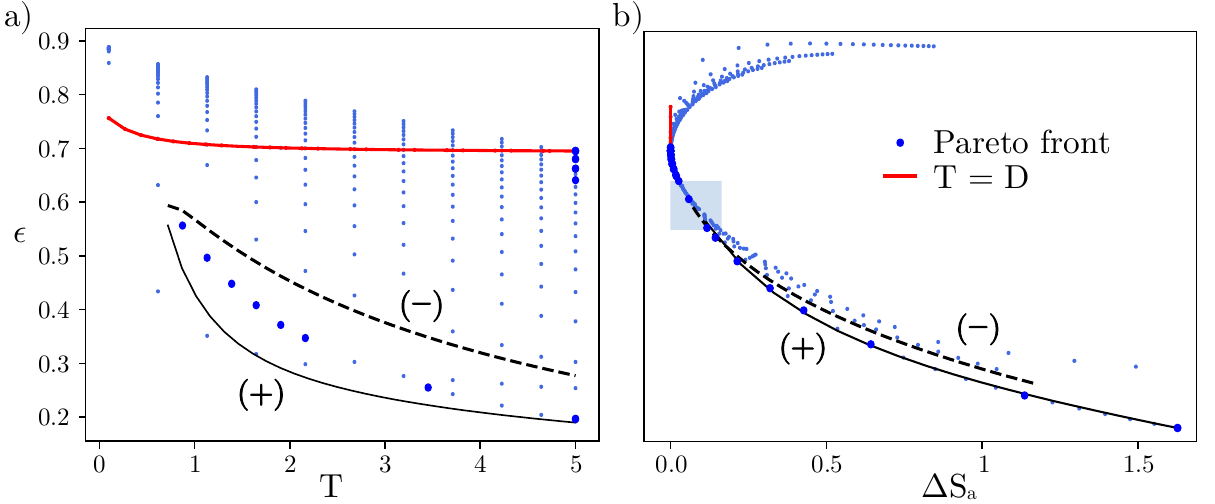}
    \caption{Adaptation error as a function of
      (a)~$T$ and (b)~$\Delta S_{\rm a}$ in
      the chemical reaction network from Fig.~\ref{fig:chemical_dynamics}a. Blue dots represent
      different values of $T, D \in [0, 5]$, the solid and dashed black
      lines the $(\pm)$ curves in
      Fig.~\ref{fig:chemical_dynamics}b, the red line corresponds to
      $T = D$, and the blue curve is the Pareto front. The blue square indicates the region where the Pareto front becomes concave. The initial condition is $p_{R}(0) = 0.9$, $p_{R_L}(0) = 0.1$
      and $p_{R_L}(0) = 0$. Other parameters are the same as in
      Fig.~\ref{fig:chemical_dynamics}. }
    \label{fig:Pareto_chemical}
\end{figure*}

The adaptation error $\epsilon(t = 1)$ is shown in Fig.~\ref{fig:Pareto_chemical} as a
function of the concentration of $T$ and $\Delta
S_{\rm a}$ for several values of $T, D \in [0,5]$.
Protein-synthesis process are
intermittent \cite{Depken2013Oct}; we thus consider the initial
concentrations to be $p_R(0)= 0.9$, $p_{R_L}(0) = 0.1$ and $p_{R_L}^*(0) = 0$, i.e.\ the absence of the complex prior to
the arrival of the ligand. 
The black lines correspond to the curves $(\pm)$ in
Fig.~\ref{fig:chemical_dynamics}b and the red line to
points with $T = D$. For low $T$, $\epsilon(1)$ decays fastest
close to
the detailed-balance case (red curve). In this regime, the adaptation is accelerated due to
an increasing $L$ at $\Delta \mu = 0$. Upon increasing $T$, the
detailed-balance curve saturates and there is not further reduction
of the adaptation error. In this regime of $T$, the adaptation error decreases fastest along
the $(+)$ line, in agreement with the relaxation time in
Fig.~\ref{fig:chemical_dynamics}b. For this set of parameters $\Delta \mu > 0$
grows, giving rise to strong clockwise currents in the network. 

The parameter set  $T = D$ satisfies detailed balance and therefore the
red line falls into the low-entropy-production region (see
Fig.~\ref{fig:Pareto_chemical}b). The curves $(\pm)$ are quite similar;
along the $(+)$ line the adaptation
error becomes reduced maximally, whereas along the $(-)$ line $\Delta
\mu$ is smaller (see Fig.~\ref{fig:chemical_dynamics}b).
 
In the convex domain the Pareto front is determined as before, by finding the values of $T$ and $D$ that minimize 
\begin{equation}
    \mathcal{P}\!=\!\inf_{T, D|\alpha}[\alpha \epsilon (T, D; t) +
      (1-\alpha)\Delta S (T, D; t)],\,\,\forall \alpha \in (0,1),
    \label{Pareto_netw}
\end{equation}
i.e.\, each point in the Pareto front corresponds to a single value of
$\alpha$.  The Pareto front obtained via Eq.~\eqref{Pareto_netw} is
shown in Fig.~\ref{fig:Pareto_chemical}b (blue line). It contains two
convex sections. In the region indicated by a blue square, we observe a concave section, whose 
consequences are echoed in Fig.~\ref{fig:Pareto_chemical}a, where we
observe that the convex sections of the Pareto front are in fact
completely disjoint. One section coincides with the detailed-balance regime $T = D$, the
other with the $(+)$ line. This reveals the existence of two distinct
optimization mechanisms that we have already hypothesized based on
intuition. In the concave section, these mechanism are said to \textit{coexist}\cite{Horowitz, Seoane2016May}. Along the detailed-balance line $T = D$, the system only
poorly improves $\epsilon$, but with the advantage of zero adiabatic dissipation, since the adaptation condition Eq.~\eqref{eq:L_constraint} can always be met for growing $L$. Along
the $(+)$ line, the system is very efficient in reducing $\epsilon$,
albeit at the expense of producing increasing amounts of entropy. The
relation between the Pareto front and the region with complex eigenvalues,
is similar to that in Fig.~\ref{fig:Pareto_fronts}. 
 
\section{Discussion and conclusion}

We investigated the adaptation ability of two biologically relevant
model systems, the cellular thermostat and a minimal ATP-driven
receptor-ligand reaction network. In the context of
the thermostat, we considered how the synthesis and back-action of
heat/cold-shock proteins regulates the effective temperature experienced by
proteins under different environmental conditions (a single
temperature shock, time-periodic temperature cycles, and stochastic
temperature shocks).  In the context of
receptor-ligand reactions we focused on how the relaxation speed and terminal
state of the formed complex upon the change in ligand concentration
are affected by the coupling to an
ATP-hydrolysis reaction that is in turn controlled by the difference between ATP and ADP concentrations. 
 
We analyzed the adaptation accuracy of both systems and determined the
respective Pareto fronts, i.e. optimal solutions to the trade-off
between adaptation accuracy and dissipation. 
These solutions do \emph{not} optimize adaptation, but instead
determine a compromise between the adaptation error and entropy
production.
Their structure depends on the particular external disruption, and may
contain one or more convex sections along the front. Each convex
section is associated with a particular optimization mechanism, while
in the concave sections Pareto optimality corresponds to the
coexistence of distinct mechanisms. 

\color{black}
The ideas developed in this work should apply rather generally to
noisy systems with underlying Markov dynamics, and not exclusively to
biological systems. However, it is in biology where we expect the most
relevant applications, e.g.\  in the characterization of effects of
evolution and the role of dissipation in evolutionary optimization.

In order to connect to experiments, we consider recent studies on
\textit{unicellular growth} mechanisms---the set of processes employed
by bacteria to increase their own biomass.
These processes are regulated by receptor-ligand networks similar to that studied in
Sec.~\ref{sec:chemical} \cite{Phillips2020Sep}, and the possible
outcomes are constrained by the underlying structure of the
chemical networks and protein structures \cite{Scott2023May}.
In
\cite{Cossetto2024Aug, Calabrese2021Nov} the authors propose that the bacterial
growth is limited by the entropy dissipation. In particular, they observe
a linear relationship between growth and free-energy consumption among a
broad range of unicellular species and metabolic types, with
dissipation rates between 10 and $10^4$ kJ/mol and a dissipation-growth
constant of 600 kJ dissipated per mol carbon grown.
Our results may
explain this feature in terms of a trade-off between growth control
and dissipation. We argue that the linear relation observed in
\cite{Cossetto2024Aug} is a Pareto front, analogous to that analyzed
in our work. If this holds, all the analyzed species  have evolved
towards the Pareto front following a \textit{central} underlying optimization
mechanism, which would explain the unique dissipation-growth constant
holding over a broad range of scales. 

Conversely, we have shown that the response of some systems
changes abruptly upon minor environmental changes. For example, minor
changes in temperature induce starkly different responses in different organisms
\cite{Bettencourt2002Sep,Hoffmann1999Dec, Buckley2001Oct}. It is
conceivable that these marked changes may be explained by the
splitting of Pareto fronts due to the coexistence of optimization
mechanisms observed for example in Figs.~\ref{fig:harmonic} and
\ref{fig:Pareto_chemical}. This hypothesis could be tested with studies
such as in Ref.~\cite{Cossetto2024Aug}. We believe that our
results capture the essential aspects of the trade-off
between adaptation and dissipation, and hope for our simple examples
to inspire and guide future experiments in this
direction. 
\color{black}

\section{Acknowledgments}
Financial support from the Alexander von
Humboldt foundation (postdoctoral fellowship to JTB)
as well as the 
German Research Foundation (DFG) through the Heisenberg Program (grant
GO 2762/4-1 to AG) and European Research Council (ERC) under the
European Union’s Horizon Europe research and innovation program (grant
agreement No 101086182 to AG) is gratefully acknowledged.

\appendix
\setcounter{equation}{0}
\setcounter{figure}{0}
\renewcommand{\thefigure}{A\arabic{figure}}
\renewcommand{\theequation}{A\arabic{equation}}
\section{cellular thermostat -- dynamics and dissipation}\label{app:steady}

\subsection{Steady-state solutions}

We write the Lyapunov equation, Eq.~\eqref{eq:Lyapunov} in components
\begin{equation}
    \begin{split}
        \frac{d\langle x_t^2 \rangle}{dt} &= -2k\langle x_t^2 \rangle - 2a\langle x_ty_t\rangle + \sigma_x^2,\\
        \frac{ d\langle x_ty_t\rangle}{dt} &= -(\gamma+k)\langle x_ty_t\rangle + \gamma\langle x_t^2 \rangle - a\langle y_t^2 \rangle,\\
        \frac{ d\langle y_t^2 \rangle}{dt} &= -2\gamma\langle y_t^2 \rangle + 2\gamma \langle x_ty_t\rangle + \gamma^2\sigma_y^2,
    \end{split}
\end{equation}
whose steady state  is given by $\frac{d}{dt} \langle x_t^2 \rangle =
\frac{d}{dt}\langle x_ty_t \rangle = \frac{d}{dt}\langle y_t^2
\rangle=0$ which yields
Eq.~\eqref{eq:Lyapunov}, i.e.\
\begin{equation}
    \langle y^2 \rangle_{\rm s} = \frac{2\langle xy^2 \rangle_{\rm s} + \gamma\sigma_y^2}{2},
\end{equation}
as well as
\begin{equation}
    \langle xy \rangle_{\rm s} = \frac{\gamma}{a + k + \gamma}\left(\langle x^2 \rangle_{\rm s} - \frac{a\sigma^2_y}{2}\right),
\end{equation}
and finally also 
\begin{equation}
    2k\langle x^2 \rangle_{\rm s} = \sigma_x^2 - \frac{2a\gamma}{a + k + \gamma}\left(\langle x^2 \rangle_{\rm s} - \frac{a\sigma^2_y}{2}\right).
\end{equation}
The steady-state values depend on $\gamma$, $a$, $\sigma_x$ and
$\sigma_y$. The adapted state satisfies the condition $\langle x^2
\rangle_{\rm s} = \sigma_x^2/2k$, implying the constraint $\sigma_y^2
= \sigma_x^2/ka$. Under this constraint we in turn find $\langle xy
\rangle_{\rm s} = 0$ and $\langle y^2 \rangle_{\rm s} = \gamma
\sigma_y^2/2$. 

\subsection{Adiabatic entropy production}
\label{sec:adiabatic_appendix}

The adiabatic component of the total entropy production rate,
Eq.~\eqref{eq:adiabatic_ent_definition}, simplifies  for Gaussian
states as follows. An instantaneous Gaussian state is defined by the probability density
\begin{equation}
    p_t(\textbf{x}) = \mathcal{N}_t\ee{-\textbf{x}^T\Sigma_t^{-1}\textbf{x}/2},
\end{equation}
where $\Sigma$ is the time-dependent correlation matrix and $\mathcal{N}_t$ is a normalization factor. In the steady state, once we impose Eq.~\eqref{eq:condition}, we have
\begin{equation}
    p_{\rm s}(\textbf{x})\equiv p_{t\to\infty}(\textbf{x})= \mathcal{N}_{\rm s}\exp\left(-\frac{kx^2}{\sigma_x^2}-\frac{y^2}{\gamma\sigma_y^2}\right),
\end{equation}
where from follows the steady-state current
\begin{equation}
    \textbf{J}_{\rm s}(\textbf{x}) = -\left(
    \begin{matrix}
        0 & -a \\
        \gamma & 0\\
    \end{matrix}
    \right)\textbf{x}\rho_{\rm s},
\end{equation}
and finally from Eq.~\eqref{eq:adiabatic_ent_definition} also the adiabatic entropy production in Eq.~\eqref{eq:adiabatic_ent_definition}
\begin{equation}
    \dot S_{\rm a}(t) = \frac{2a}{\sigma_x^2}(a\langle y_t^2\rangle + k\langle x_t^2\rangle).
\end{equation}

\subsection{Steady state under stochastic temperature shocks} \label{app:perturbation}

Here we analyze the cellular  thermostat under
continuous-time stochastic temperature shocks, in the limit of large
and small shock (i.e.\ resetting) rate $\Gamma$. Recall that the extended Fokker-Planck equation reads
\begin{equation}
    \partial_t p_t(\textbf{x}) = \nabla\cdot \left[A\textbf{x}p_t(\textbf{x}) \!+\! D\nabla p_t(\textbf{x})\right] + \Gamma\left[p_{\rm res}(\textbf{x})\!-\!p_t(\textbf{x})\right].
\end{equation}

\subsubsection{Small-$\Gamma$ limit}

For small $\Gamma$ we can expand the solution perturbatively in orders of $\Gamma$
\begin{equation}
    p_t(\textbf{x}) = \Gamma^0 p_t^{(0)}(\textbf{x}) + \Gamma p_t^{(1)}(\textbf{x}) + \Gamma^2p_t^{(2)}(\textbf{x}) + ...
\end{equation}
Inserting back to Fokker-Planck equation and collecting matching  orders of $\Gamma$ we obtain the hierarchy of equations
\begin{align}
\partial_t p_t^{(0)} \!&= \! \nabla\cdot\! \left[A\textbf{x}p_t^{(0)}(\textbf{x}) + D\nabla p_t^{(0)}\textbf{x})\right]
        \\
    \partial_t p_t^{(1)}  \!&= \!  \nabla\cdot\!
    \left[A\textbf{x}p_t^{(1)}(\textbf{x}) + D\nabla
      p_t^{(1)}(\textbf{x})\right] \!+\! \left[p_{\rm
        res}(\textbf{x})\!-\! p_t^{(0)}(\textbf{x})\right]\nonumber\\
     \partial_t p_t^{(2)}  \!&= \!  \nabla\cdot\! \left[A\textbf{x}p_t^{(2)}(\textbf{x}) + D\nabla p_t^{(2)}(\textbf{x})\right] \!-\! p_t^{(1)}(\textbf{x})\nonumber\\
     \partial_t p_t^{(3)}  \!&= \!  \nabla\cdot\! \left[A\textbf{x}p_t^{(3)}(\textbf{x}) + D\nabla p_t^{(3)}(\textbf{x})\right] \!- \!p_t^{(2)}(\textbf{x})\nonumber\\
        \vdots \nonumber\\
     \partial_t p_t^{(n)}  \!&= \!  \nabla\cdot\! \left[A\textbf{x}p_t^{(n)}(\textbf{x}) + D\nabla p_t^{(n)}(\textbf{x})\right]\! -\! p_t^{(n-1)}(\textbf{x}).\nonumber
\end{align}
If the initial distribution is Gaussian we can evaluate the covariance
elements to any order $n$ in $\Gamma$, $\textbf{v}_t^{(n)} = (\langle x_t^2\rangle^{(n)},\langle x_ty_t\rangle^{(n)},\langle y_t^2\rangle^{(n)})$ as
\begin{align}
    \dot{\textbf{v}}_t^{(0)} &= M\textbf{v}_t^{(0)}  +  \textbf{d}\\
    \dot{\textbf{v}}_t^{(1)} &= M\textbf{v}_t^{(1)} + \textbf{d} + \textbf{v}_{\rm res} - \textbf{v}_t^{(0)} \nonumber\\
    \dot{\textbf{v}}_t^{(2)} &= M\textbf{v}_t^{(2)} + \textbf{d} - \textbf{v}_t^{(1)}  \nonumber\\
    \vdots \nonumber\\
    \dot{\textbf{v}}_t^{(n)} &= M\textbf{v}_t^{(n)} + \textbf{d} - \textbf{v}_t^{(n-1)}, \nonumber
\end{align}
where $\textbf{d} \equiv (\sigma_x^2, 0,
\gamma^2\sigma_y^2)^T$. 
In the steady state, we can solve the system of algebraic equations
$\dot{\textbf{v}}_t^{(n)} = 0$ for all $n$
\begin{equation}
\begin{split}
        \textbf{v}^{(0)} =& -M^{-1}\textbf{d}
        \\
        \textbf{v}^{(n)} =& -M^{-n}\textbf{v}_{\rm res} -\sum_{i = 0}^n M^{-i-1}\textbf{d}\nonumber\\
    =&-M^{-n}\textbf{v}_{\rm res}-M^{-1}(\mathbbm{I}-M^{-1})^{-1}(\mathbbm{I}-M^{-n-1})\textbf{d},
\end{split}
\end{equation}
and evaluate the covariance vector $\textbf{v}_{\rm NESS} =
\textbf{v}_0 + \textbf{v}_1\Gamma + ...$ yielding
\begin{align}
    \textbf{v}_{\rm NESS}=& -M^{-1} \textbf{d}- \sum_{n = 1}^\infty
    \Gamma^n\left(M^{-n}\textbf{v}_{\rm res} +\sum_{i = 0}^n
    M^{-i-1}\textbf{d}\right)\nonumber\\
    =&-M^{-1} \textbf{d}- \Gamma M^{-1}(\mathbbm{I}-\Gamma
      M^{-1})^{-1}\textbf{v}_{\rm res}\\
    &- \frac{\Gamma}{1-\Gamma}M^{-1}(\mathbbm{I}-M^{-1})^{-1}\textbf{d}\nonumber\\
    &+ \Gamma M^{-3}(\mathbbm{I}-M^{-1})^{-1}(\mathbbm{I}-\Gamma M^{-1})^{-1}\textbf{d},\nonumber
\end{align}
and the convergence of the sums is guaranteed by taking $\Gamma$
small enough to ensure $\lim_{n\to\infty}(\Gamma M^{-1})^n=0$. 
To the first order in $\Gamma$ we obtain 
\begin{equation}
    \textbf{v}_{\rm NESS}= \textbf{v}_{\rm s} + \Gamma
    M^{-1}(\textbf{v}_{\rm s} - \textbf{v}_{\rm res}) +\mathcal{O}(\Gamma^2),
\end{equation}
which was considered in the main part.

\subsubsection{Large-$\Gamma$ limit}

For high $\Gamma$ we set $\veps\equiv 1/\Gamma$ and expand as
\begin{equation}
    p_t(\textbf{x}) = \veps^0 \bp_t^{(0)}(\textbf{x}) + \veps \bp_t^{(1)}(\textbf{x}) + \veps^2\bp_t^{(2)}(\textbf{x}) + ...
\end{equation}
and collect orders of $\veps$
\begin{align}
\partial_t \bp_t^{(0)} \!&= \! \nabla\cdot\! \left[A\textbf{x}\bp_t^{(0)}(\textbf{x}) + D\nabla \bp_t^{(0)}\textbf{x})\right]- \bp_t^{(1)}(\textbf{x})\nonumber
        \\
    \partial_t \bp_t^{(1)}  \!&= \!  \nabla\cdot\!
    \left[A\textbf{x}\bp_t^{(1)}(\textbf{x}) + D\nabla
      \bp_t^{(1)}(\textbf{x})\right] \!+\! \left[p_{\rm
        res}(\textbf{x})\!-\! \bp_t^{(2)}(\textbf{x})\right]\nonumber\\
     \partial_t \bp_t^{(2)}  \!&= \!  \nabla\cdot\! \left[A\textbf{x}\bp_t^{(2)}(\textbf{x}) + D\nabla \bp_t^{(2)}(\textbf{x})\right] \!-\! \bp_t^{(3)}(\textbf{x})\nonumber\\
     \partial_t \bp_t^{(3)}  \!&= \!  \nabla\cdot\! \left[A\textbf{x}\bp_t^{(3)}(\textbf{x}) + D\nabla \bp_t^{(3)}(\textbf{x})\right] \!- \!\bp_t^{(4)}(\textbf{x})\nonumber\\
        \vdots \nonumber\\
     \partial_t \bp_t^{(n)}  \!&= \!  \nabla\cdot\! \left[A\textbf{x}\bp_t^{(n)}(\textbf{x}) + D\nabla \bp_t^{(n)}(\textbf{x})\right]\! -\! \bp_t^{(n+1)}(\textbf{x}).\nonumber
\end{align}
Again, for a Gaussian initial condition we find 
\begin{align}
    \dot{\bv}_t^{(0)} &= -M\bv_t^{(0)}  +  \textbf{d}-\bv_t^{(1)}\nonumber\\
    \dot{\bv}_t^{(1)} &= -M\bv_t^{(1)} + \textbf{d} + \textbf{v}_{\rm res} - \bv_t^{(2)} \nonumber\\
    \dot{\bv}_t^{(2)} &= -M\bv_t^{(2)} + \textbf{d} - \bv_t^{(3)}\\
    \vdots \nonumber\\
    \dot{\bv}_t^{(n)} &= -M\bv_t^{(n)} + \textbf{d} - \bv_t^{(n+1)}, \nonumber
\end{align}
and the steady-state follows straightforwardly, once we identify
$\textbf{v}_0 = \textbf{v}_{\rm res}$ (i.e.\ the limit $\Gamma\to\infty$) yielding,
\begin{align}
        \bv^{(0)} =& \textbf{v}_{\rm res},
        \\
        \bv^{(1)} =& -M \textbf{v}_{\rm res} + \textbf{d}\nonumber
        \\\nonumber
        \vdots
\end{align}
To first order in $\Gamma^{-1}$ we thus obtain
\begin{equation}
    \textbf{v}_{\rm NESS} = \textbf{v}_{\rm res} + M(\textbf{v}_{\rm
      res}-\textbf{v}_{\rm s})/\Gamma +\mathcal{O}(\Gamma^{-2}),
\end{equation}
which is used in the main part.
\begin{figure*}
    \centering
\includegraphics[width = \linewidth]{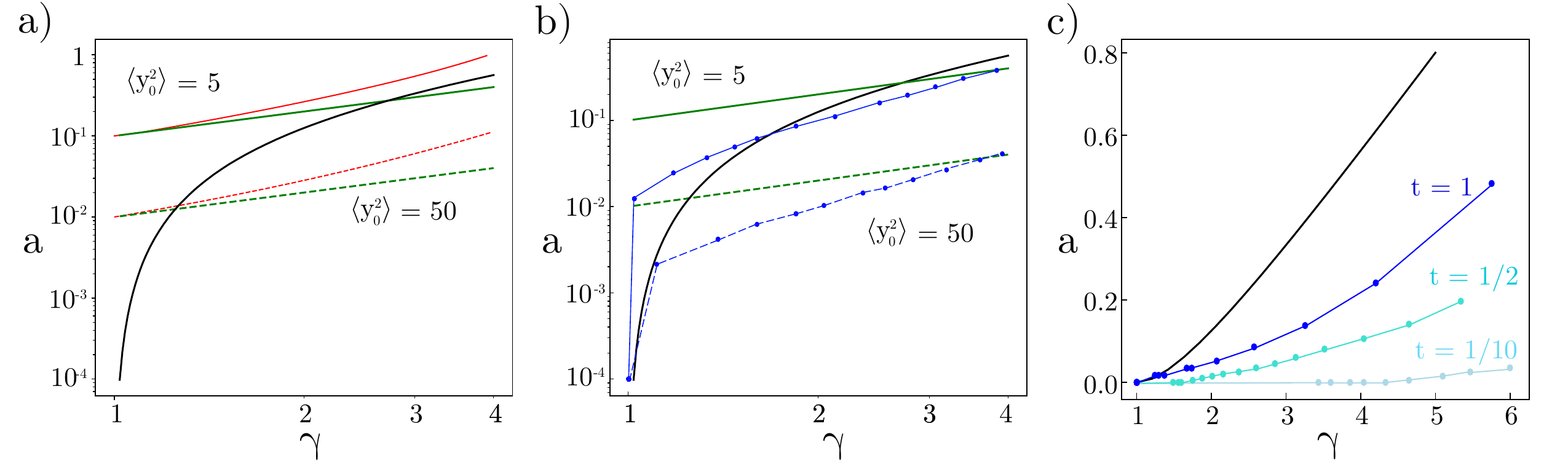}
    \caption{(a)~Minimal adaptation error (red lines)
      and (b) Pareto front (blue) during relaxation of the
      cellular thermostat at time $t = 1$ for two different
      values of $\langle y^2_0\rangle$ shown in the $\gamma,\alpha$
      space.~In both (a) and (b) he black line represents $a = (k-\gamma)^2 /4\gamma$
      curve and the green one corresponds to $a = \gamma
      \sigma_x^2/2k\langle y^2_0\rangle$ [see
        Eq.~\eqref{eq:curva_verde}]. (c)~Pareto fronts
        in the $\gamma,\alpha$ space at different times during
        relaxation of the cellular thermostat; the black line
        denotes $a = (k-\gamma)^2 /4\gamma$. The remaining parameters are
      taken as in Fig.~\eqref{fig:Eigenvalues} }
    \label{fig:initial_condition}
\end{figure*}

\setcounter{equation}{0}
\setcounter{figure}{0}
\renewcommand{\thefigure}{B\arabic{figure}}
\renewcommand{\theequation}{B\arabic{equation}}

\section{Additional information on the cellular thermostat}\label{app:additional_plots}
Here we provide additional data supporting the results from the main
text. We consider the cellular thermostat during the relaxation
process in Fig.~\ref{fig:Eigenvalues}c.

Results for the minimal adaptation error (corresponding to the red
lines shown in Fig.~\ref{fig:Pareto_fronts}) for two different values
of $\langle y_0^2\rangle$ are shown in Fig.~\ref{fig:initial_condition}a.
The green lines correspond to $a = \gamma \sigma_x^2/2k\langle
y_0^2\rangle$ and  contain information about the initial state of
$y$.
In Fig.~\ref{fig:initial_condition}b we show Pareto fronts
corresponding to the blue lines in Fig.~\ref{fig:Pareto_fronts}. We
observe how the blue and red lines vary with $\langle y_0^2\rangle$.
The black line in Fig.~\ref{fig:initial_condition} corresponds to  $a
= (k-\gamma)^2 /4\gamma$ and coincides with the black line in
Fig.~\ref{fig:Eigenvalues}b and, notably, does not depend on $\langle
y_0^2\rangle$.

In Fig.~\ref{fig:initial_condition}c we show Pareto fronts in the
$\gamma, \alpha$ space at
several times during  relaxation. The black line corresponds again to $a =
(k-\gamma)^2 /4\gamma$. At early times, the Pareto front remains close
to $a = 0$ because the feedback mechanism has no time to reduce the
adaptation error $\epsilon(t=0.1)$, and the unique optimal solution
corresponds to reducing dissipation. At larger times, the feedback
system becomes more efficient, and the Pareto front moves closer to
the black line. Increasing $t$ implies a reduction of the adaptation error.

\bibliography{biblio}

\begin{thebibliography}{60}%
\makeatletter
\providecommand \@ifxundefined [1]{%
 \@ifx{#1\undefined}
}%
\providecommand \@ifnum [1]{%
 \ifnum #1\expandafter \@firstoftwo
 \else \expandafter \@secondoftwo
 \fi
}%
\providecommand \@ifx [1]{%
 \ifx #1\expandafter \@firstoftwo
 \else \expandafter \@secondoftwo
 \fi
}%
\providecommand \natexlab [1]{#1}%
\providecommand \enquote  [1]{``#1''}%
\providecommand \bibnamefont  [1]{#1}%
\providecommand \bibfnamefont [1]{#1}%
\providecommand \citenamefont [1]{#1}%
\providecommand \href@noop [0]{\@secondoftwo}%
\providecommand \href [0]{\begingroup \@sanitize@url \@href}%
\providecommand \@href[1]{\@@startlink{#1}\@@href}%
\providecommand \@@href[1]{\endgroup#1\@@endlink}%
\providecommand \@sanitize@url [0]{\catcode `\\12\catcode `\$12\catcode
  `\&12\catcode `\#12\catcode `\^12\catcode `\_12\catcode `\%12\relax}%
\providecommand \@@startlink[1]{}%
\providecommand \@@endlink[0]{}%
\providecommand \url  [0]{\begingroup\@sanitize@url \@url }%
\providecommand \@url [1]{\endgroup\@href {#1}{\urlprefix }}%
\providecommand \urlprefix  [0]{URL }%
\providecommand \Eprint [0]{\href }%
\providecommand \doibase [0]{https://doi.org/}%
\providecommand \selectlanguage [0]{\@gobble}%
\providecommand \bibinfo  [0]{\@secondoftwo}%
\providecommand \bibfield  [0]{\@secondoftwo}%
\providecommand \translation [1]{[#1]}%
\providecommand \BibitemOpen [0]{}%
\providecommand \bibitemStop [0]{}%
\providecommand \bibitemNoStop [0]{.\EOS\space}%
\providecommand \EOS [0]{\spacefactor3000\relax}%
\providecommand \BibitemShut  [1]{\csname bibitem#1\endcsname}%
\let\auto@bib@innerbib\@empty
%</preamble>
\bibitem [{\citenamefont {Astumian}\ and\ \citenamefont
  {H{\ifmmode\ddot{a}\else\"{a}\fi}nggi}(2002)}]{Astumian2002Nov}%
  \BibitemOpen
  \bibfield  {author} {\bibinfo {author} {\bibfnamefont {R.~D.}\ \bibnamefont
  {Astumian}}\ and\ \bibinfo {author} {\bibfnamefont {P.}~\bibnamefont
  {H{\ifmmode\ddot{a}\else\"{a}\fi}nggi}},\ }\bibfield  {title} {\bibinfo
  {title} {{Brownian Motors}},\ }\href {https://doi.org/10.1063/1.1535005}
  {\bibfield  {journal} {\bibinfo  {journal} {Phys. Today}\ }\textbf {\bibinfo
  {volume} {55}},\ \bibinfo {pages} {33} (\bibinfo {year} {2002})}\BibitemShut
  {NoStop}%
\bibitem [{\citenamefont {Astumian}(2019)}]{Astumian2019Aug}%
  \BibitemOpen
  \bibfield  {author} {\bibinfo {author} {\bibfnamefont {R.~D.}\ \bibnamefont
  {Astumian}},\ }\bibfield  {title} {\bibinfo {title} {{Kinetic asymmetry
  allows macromolecular catalysts to drive an information ratchet}},\ }\href
  {https://doi.org/10.1038/s41467-019-11402-7} {\bibfield  {journal} {\bibinfo
  {journal} {Nat. Commun.}\ }\textbf {\bibinfo {volume} {10}},\ \bibinfo
  {pages} {1} (\bibinfo {year} {2019})}\BibitemShut {NoStop}%
\bibitem [{\citenamefont {Amano}\ \emph {et~al.}(2022)\citenamefont {Amano},
  \citenamefont {Esposito}, \citenamefont {Kreidt}, \citenamefont {Leigh},
  \citenamefont {Penocchio},\ and\ \citenamefont {Roberts}}]{Amano2022May}%
  \BibitemOpen
  \bibfield  {author} {\bibinfo {author} {\bibfnamefont {S.}~\bibnamefont
  {Amano}}, \bibinfo {author} {\bibfnamefont {M.}~\bibnamefont {Esposito}},
  \bibinfo {author} {\bibfnamefont {E.}~\bibnamefont {Kreidt}}, \bibinfo
  {author} {\bibfnamefont {D.~A.}\ \bibnamefont {Leigh}}, \bibinfo {author}
  {\bibfnamefont {E.}~\bibnamefont {Penocchio}},\ and\ \bibinfo {author}
  {\bibfnamefont {B.~M.~W.}\ \bibnamefont {Roberts}},\ }\bibfield  {title}
  {\bibinfo {title} {{Insights from an information thermodynamics analysis of a
  synthetic molecular motor}},\ }\href
  {https://doi.org/10.1038/s41557-022-00899-z} {\bibfield  {journal} {\bibinfo
  {journal} {Nat. Chem.}\ }\textbf {\bibinfo {volume} {14}},\ \bibinfo {pages}
  {530} (\bibinfo {year} {2022})}\BibitemShut {NoStop}%
\bibitem [{\citenamefont {Boeger}(2022)}]{Boeger2022Jun}%
  \BibitemOpen
  \bibfield  {author} {\bibinfo {author} {\bibfnamefont {H.}~\bibnamefont
  {Boeger}},\ }\bibfield  {title} {\bibinfo {title} {{Kinetic Proofreading}},\
  }\href {https://doi.org/10.1146/annurev-biochem-040320-103630} {\bibfield
  {journal} {\bibinfo  {journal} {Annu. Rev. Biochem.}\ ,\ \bibinfo {pages}
  {423}} (\bibinfo {year} {2022})}\BibitemShut {NoStop}%
\bibitem [{\citenamefont {Qian}(2007)}]{Qian2007May}%
  \BibitemOpen
  \bibfield  {author} {\bibinfo {author} {\bibfnamefont {H.}~\bibnamefont
  {Qian}},\ }\bibfield  {title} {\bibinfo {title} {{Phosphorylation Energy
  Hypothesis: Open Chemical Systems and Their Biological Functions}},\ }\href
  {https://doi.org/10.1146/annurev.physchem.58.032806.104550} {\bibfield
  {journal} {\bibinfo  {journal} {Annu. Rev. Phys. Chem.}\ ,\ \bibinfo {pages}
  {113}} (\bibinfo {year} {2007})}\BibitemShut {NoStop}%
\bibitem [{\citenamefont {Ma}\ \emph {et~al.}(2009)\citenamefont {Ma},
  \citenamefont {Trusina}, \citenamefont {El-Samad}, \citenamefont {Lim},\ and\
  \citenamefont {Tang}}]{Ma2009Aug}%
  \BibitemOpen
  \bibfield  {author} {\bibinfo {author} {\bibfnamefont {W.}~\bibnamefont
  {Ma}}, \bibinfo {author} {\bibfnamefont {A.}~\bibnamefont {Trusina}},
  \bibinfo {author} {\bibfnamefont {H.}~\bibnamefont {El-Samad}}, \bibinfo
  {author} {\bibfnamefont {W.~A.}\ \bibnamefont {Lim}},\ and\ \bibinfo {author}
  {\bibfnamefont {C.}~\bibnamefont {Tang}},\ }\bibfield  {title} {\bibinfo
  {title} {{Defining Network Topologies that Can Achieve Biochemical
  Adaptation}},\ }\href {https://doi.org/10.1016/j.cell.2009.06.013} {\bibfield
   {journal} {\bibinfo  {journal} {Cell}\ }\textbf {\bibinfo {volume} {138}},\
  \bibinfo {pages} {760} (\bibinfo {year} {2009})}\BibitemShut {NoStop}%
\bibitem [{\citenamefont {Smirnakis}\ \emph {et~al.}(1997)\citenamefont
  {Smirnakis}, \citenamefont {Berry}, \citenamefont {Warland}, \citenamefont
  {Bialek},\ and\ \citenamefont {Meister}}]{Smirnakis1997Mar}%
  \BibitemOpen
  \bibfield  {author} {\bibinfo {author} {\bibfnamefont {S.~M.}\ \bibnamefont
  {Smirnakis}}, \bibinfo {author} {\bibfnamefont {M.~J.}\ \bibnamefont
  {Berry}}, \bibinfo {author} {\bibfnamefont {D.~K.}\ \bibnamefont {Warland}},
  \bibinfo {author} {\bibfnamefont {W.}~\bibnamefont {Bialek}},\ and\ \bibinfo
  {author} {\bibfnamefont {M.}~\bibnamefont {Meister}},\ }\bibfield  {title}
  {\bibinfo {title} {{Adaptation of retinal processing to image contrast and
  spatial scale}},\ }\href {https://doi.org/10.1038/386069a0} {\bibfield
  {journal} {\bibinfo  {journal} {Nature}\ }\textbf {\bibinfo {volume} {386}},\
  \bibinfo {pages} {69} (\bibinfo {year} {1997})}\BibitemShut {NoStop}%
\bibitem [{\citenamefont {Hazelbauer}\ \emph {et~al.}(2008)\citenamefont
  {Hazelbauer}, \citenamefont {Falke},\ and\ \citenamefont
  {Parkinson}}]{Hazelbauer2008Jan}%
  \BibitemOpen
  \bibfield  {author} {\bibinfo {author} {\bibfnamefont {G.~L.}\ \bibnamefont
  {Hazelbauer}}, \bibinfo {author} {\bibfnamefont {J.~J.}\ \bibnamefont
  {Falke}},\ and\ \bibinfo {author} {\bibfnamefont {J.~S.}\ \bibnamefont
  {Parkinson}},\ }\bibfield  {title} {\bibinfo {title} {{Bacterial
  chemoreceptors: high-performance signaling in networked arrays}},\ }\href
  {https://doi.org/10.1016/j.tibs.2007.09.014} {\bibfield  {journal} {\bibinfo
  {journal} {Trends Biochem. Sci.}\ }\textbf {\bibinfo {volume} {33}},\
  \bibinfo {pages} {9} (\bibinfo {year} {2008})}\BibitemShut {NoStop}%
\bibitem [{\citenamefont {Thieringer}\ \emph {et~al.}(1998)\citenamefont
  {Thieringer}, \citenamefont {Jones},\ and\ \citenamefont
  {Inouye}}]{Thieringer1998Jan}%
  \BibitemOpen
  \bibfield  {author} {\bibinfo {author} {\bibfnamefont {H.~A.}\ \bibnamefont
  {Thieringer}}, \bibinfo {author} {\bibfnamefont {P.~G.}\ \bibnamefont
  {Jones}},\ and\ \bibinfo {author} {\bibfnamefont {M.}~\bibnamefont
  {Inouye}},\ }\bibfield  {title} {\bibinfo {title} {{Cold shock and
  adaptation}},\ }\href
  {https://doi.org/10.1002/(SICI)1521-1878(199801)20:1<49::AID-BIES8>3.0.CO;2-N}
  {\bibfield  {journal} {\bibinfo  {journal} {BioEssays}\ }\textbf {\bibinfo
  {volume} {20}},\ \bibinfo {pages} {49} (\bibinfo {year} {1998})}\BibitemShut
  {NoStop}%
\bibitem [{\citenamefont {Etchegaray}\ \emph {et~al.}(1996)\citenamefont
  {Etchegaray}, \citenamefont {Jones},\ and\ \citenamefont
  {Inouye}}]{Etchegaray1996Feb}%
  \BibitemOpen
  \bibfield  {author} {\bibinfo {author} {\bibfnamefont {J.-P.}\ \bibnamefont
  {Etchegaray}}, \bibinfo {author} {\bibfnamefont {P.~G.}\ \bibnamefont
  {Jones}},\ and\ \bibinfo {author} {\bibfnamefont {M.}~\bibnamefont
  {Inouye}},\ }\bibfield  {title} {\bibinfo {title} {{Differential
  thermoregulation of two highly homologous cold-shock genes, and , of}},\
  }\href {https://doi.org/10.1046/j.1365-2443.1996.d01-231.x} {\bibfield
  {journal} {\bibinfo  {journal} {Genes Cells}\ }\textbf {\bibinfo {volume}
  {1}},\ \bibinfo {pages} {171} (\bibinfo {year} {1996})}\BibitemShut {NoStop}%
\bibitem [{\citenamefont {Buckley}\ \emph {et~al.}(2001)\citenamefont
  {Buckley}, \citenamefont {Owen},\ and\ \citenamefont
  {Hofmann}}]{Buckley2001Oct}%
  \BibitemOpen
  \bibfield  {author} {\bibinfo {author} {\bibfnamefont {B.~A.}\ \bibnamefont
  {Buckley}}, \bibinfo {author} {\bibfnamefont {M.-E.}\ \bibnamefont {Owen}},\
  and\ \bibinfo {author} {\bibfnamefont {G.~E.}\ \bibnamefont {Hofmann}},\
  }\bibfield  {title} {\bibinfo {title} {{Adjusting the thermostat: the
  threshold induction temperature for the heat-shock response in intertidal
  mussels (genus Mytilus) changes as a function of thermal history}},\ }\href
  {https://doi.org/10.1242/jeb.204.20.3571} {\bibfield  {journal} {\bibinfo
  {journal} {J. Exp. Biol.}\ }\textbf {\bibinfo {volume} {204}},\ \bibinfo
  {pages} {3571} (\bibinfo {year} {2001})}\BibitemShut {NoStop}%
\bibitem [{\citenamefont {Sriram}\ \emph {et~al.}(2012)\citenamefont {Sriram},
  \citenamefont {Rodriguez-Fernandez},\ and\ \citenamefont
  {Iii}}]{Sriram2012Aug}%
  \BibitemOpen
  \bibfield  {author} {\bibinfo {author} {\bibfnamefont {K.}~\bibnamefont
  {Sriram}}, \bibinfo {author} {\bibfnamefont {M.}~\bibnamefont
  {Rodriguez-Fernandez}},\ and\ \bibinfo {author} {\bibfnamefont {F.~J.~D.}\
  \bibnamefont {Iii}},\ }\bibfield  {title} {\bibinfo {title} {{A Detailed
  Modular Analysis of Heat-Shock Protein Dynamics under Acute and Chronic
  Stress and Its Implication in Anxiety Disorders}},\ }\href
  {https://doi.org/10.1371/journal.pone.0042958} {\bibfield  {journal}
  {\bibinfo  {journal} {PLoS One}\ }\textbf {\bibinfo {volume} {7}},\ \bibinfo
  {pages} {e42958} (\bibinfo {year} {2012})}\BibitemShut {NoStop}%
\bibitem [{\citenamefont {Siv{\ifmmode\acute{e}\else\'{e}\fi}ry}\ \emph
  {et~al.}(2016)\citenamefont {Siv{\ifmmode\acute{e}\else\'{e}\fi}ry},
  \citenamefont {Courtade},\ and\ \citenamefont {Thommen}}]{Sivery2016Dec}%
  \BibitemOpen
  \bibfield  {author} {\bibinfo {author} {\bibfnamefont {A.}~\bibnamefont
  {Siv{\ifmmode\acute{e}\else\'{e}\fi}ry}}, \bibinfo {author} {\bibfnamefont
  {E.}~\bibnamefont {Courtade}},\ and\ \bibinfo {author} {\bibfnamefont
  {Q.}~\bibnamefont {Thommen}},\ }\bibfield  {title} {\bibinfo {title} {{A
  minimal titration model of the mammalian dynamical heat shock response}},\
  }\href {https://doi.org/10.1088/1478-3975/13/6/066008} {\bibfield  {journal}
  {\bibinfo  {journal} {Phys. Biol.}\ }\textbf {\bibinfo {volume} {13}},\
  \bibinfo {pages} {066008} (\bibinfo {year} {2016})}\BibitemShut {NoStop}%
\bibitem [{\citenamefont {Schmiedl}\ and\ \citenamefont
  {Seifert}(2007)}]{Schmiedl2007Mar}%
  \BibitemOpen
  \bibfield  {author} {\bibinfo {author} {\bibfnamefont {T.}~\bibnamefont
  {Schmiedl}}\ and\ \bibinfo {author} {\bibfnamefont {U.}~\bibnamefont
  {Seifert}},\ }\bibfield  {title} {\bibinfo {title} {{Optimal Finite-Time
  Processes In Stochastic Thermodynamics}},\ }\href
  {https://doi.org/10.1103/PhysRevLett.98.108301} {\bibfield  {journal}
  {\bibinfo  {journal} {Phys. Rev. Lett.}\ }\textbf {\bibinfo {volume} {98}},\
  \bibinfo {pages} {108301} (\bibinfo {year} {2007})}\BibitemShut {NoStop}%
\bibitem [{\citenamefont {Falasco}\ \emph {et~al.}(2022)\citenamefont
  {Falasco}, \citenamefont {Esposito},\ and\ \citenamefont
  {Delvenne}}]{Falasco2022Feb}%
  \BibitemOpen
  \bibfield  {author} {\bibinfo {author} {\bibfnamefont {G.}~\bibnamefont
  {Falasco}}, \bibinfo {author} {\bibfnamefont {M.}~\bibnamefont {Esposito}},\
  and\ \bibinfo {author} {\bibfnamefont {J.-C.}\ \bibnamefont {Delvenne}},\
  }\bibfield  {title} {\bibinfo {title} {{Beyond thermodynamic uncertainty
  relations: nonlinear response, error-dissipation trade-offs, and speed
  limits}},\ }\href {https://doi.org/10.1088/1751-8121/ac52e2} {\bibfield
  {journal} {\bibinfo  {journal} {J. Phys. A: Math. Theor.}\ }\textbf {\bibinfo
  {volume} {55}},\ \bibinfo {pages} {124002} (\bibinfo {year}
  {2022})}\BibitemShut {NoStop}%
\bibitem [{\citenamefont {Gu{\ifmmode\acute{e}\else\'{e}\fi}ry-Odelin}\ \emph
  {et~al.}(2023)\citenamefont {Gu{\ifmmode\acute{e}\else\'{e}\fi}ry-Odelin},
  \citenamefont {Jarzynski}, \citenamefont {Plata}, \citenamefont {Prados},\
  and\ \citenamefont {Trizac}}]{Guery-Odelin2023Jan}%
  \BibitemOpen
  \bibfield  {author} {\bibinfo {author} {\bibfnamefont {D.}~\bibnamefont
  {Gu{\ifmmode\acute{e}\else\'{e}\fi}ry-Odelin}}, \bibinfo {author}
  {\bibfnamefont {C.}~\bibnamefont {Jarzynski}}, \bibinfo {author}
  {\bibfnamefont {C.~A.}\ \bibnamefont {Plata}}, \bibinfo {author}
  {\bibfnamefont {A.}~\bibnamefont {Prados}},\ and\ \bibinfo {author}
  {\bibfnamefont {E.}~\bibnamefont {Trizac}},\ }\bibfield  {title} {\bibinfo
  {title} {{Driving rapidly while remaining in control: classical shortcuts
  from Hamiltonian to stochastic dynamics}},\ }\href
  {https://doi.org/10.1088/1361-6633/acacad} {\bibfield  {journal} {\bibinfo
  {journal} {Rep. Prog. Phys.}\ }\textbf {\bibinfo {volume} {86}},\ \bibinfo
  {pages} {035902} (\bibinfo {year} {2023})}\BibitemShut {NoStop}%
\bibitem [{\citenamefont {Plata}\ \emph {et~al.}(2020)\citenamefont {Plata},
  \citenamefont {Gu{\ifmmode\acute{e}\else\'{e}\fi}ry-Odelin}, \citenamefont
  {Trizac},\ and\ \citenamefont {Prados}}]{Plata2020Mar}%
  \BibitemOpen
  \bibfield  {author} {\bibinfo {author} {\bibfnamefont {C.~A.}\ \bibnamefont
  {Plata}}, \bibinfo {author} {\bibfnamefont {D.}~\bibnamefont
  {Gu{\ifmmode\acute{e}\else\'{e}\fi}ry-Odelin}}, \bibinfo {author}
  {\bibfnamefont {E.}~\bibnamefont {Trizac}},\ and\ \bibinfo {author}
  {\bibfnamefont {A.}~\bibnamefont {Prados}},\ }\bibfield  {title} {\bibinfo
  {title} {{Finite-time adiabatic processes: Derivation and speed limit}},\
  }\href {https://doi.org/10.1103/PhysRevE.101.032129} {\bibfield  {journal}
  {\bibinfo  {journal} {Phys. Rev. E}\ }\textbf {\bibinfo {volume} {101}},\
  \bibinfo {pages} {032129} (\bibinfo {year} {2020})}\BibitemShut {NoStop}%
\bibitem [{\citenamefont {Solon}\ and\ \citenamefont
  {Horowitz}(2018)}]{Horowitz}%
  \BibitemOpen
  \bibfield  {author} {\bibinfo {author} {\bibfnamefont {A.~P.}\ \bibnamefont
  {Solon}}\ and\ \bibinfo {author} {\bibfnamefont {J.~M.}\ \bibnamefont
  {Horowitz}},\ }\bibfield  {title} {\bibinfo {title} {{Phase Transition in
  Protocols Minimizing Work Fluctuations}},\ }\href
  {https://doi.org/10.1103/PhysRevLett.120.180605} {\bibfield  {journal}
  {\bibinfo  {journal} {Phys. Rev. Lett.}\ }\textbf {\bibinfo {volume} {120}},\
  \bibinfo {pages} {180605} (\bibinfo {year} {2018})}\BibitemShut {NoStop}%
\bibitem [{\citenamefont {Censor}(1977)}]{Censor1977Mar}%
  \BibitemOpen
  \bibfield  {author} {\bibinfo {author} {\bibfnamefont {Y.}~\bibnamefont
  {Censor}},\ }\bibfield  {title} {\bibinfo {title} {{Pareto optimality in
  multiobjective problems}},\ }\href {https://doi.org/10.1007/BF01442131}
  {\bibfield  {journal} {\bibinfo  {journal} {Appl. Math. Optim.}\ }\textbf
  {\bibinfo {volume} {4}},\ \bibinfo {pages} {41} (\bibinfo {year}
  {1977})}\BibitemShut {NoStop}%
\bibitem [{\citenamefont {Shoval}\ \emph {et~al.}(2012)\citenamefont {Shoval},
  \citenamefont {Sheftel}, \citenamefont {Shinar}, \citenamefont {Hart},
  \citenamefont {Ramote}, \citenamefont {Mayo}, \citenamefont {Dekel},
  \citenamefont {Kavanagh},\ and\ \citenamefont {Alon}}]{Shoval2012Apr}%
  \BibitemOpen
  \bibfield  {author} {\bibinfo {author} {\bibfnamefont {O.}~\bibnamefont
  {Shoval}}, \bibinfo {author} {\bibfnamefont {H.}~\bibnamefont {Sheftel}},
  \bibinfo {author} {\bibfnamefont {G.}~\bibnamefont {Shinar}}, \bibinfo
  {author} {\bibfnamefont {Y.}~\bibnamefont {Hart}}, \bibinfo {author}
  {\bibfnamefont {O.}~\bibnamefont {Ramote}}, \bibinfo {author} {\bibfnamefont
  {A.}~\bibnamefont {Mayo}}, \bibinfo {author} {\bibfnamefont {E.}~\bibnamefont
  {Dekel}}, \bibinfo {author} {\bibfnamefont {K.}~\bibnamefont {Kavanagh}},\
  and\ \bibinfo {author} {\bibfnamefont {U.}~\bibnamefont {Alon}},\ }\bibfield
  {title} {\bibinfo {title} {{Evolutionary Trade-Offs, Pareto Optimality, and
  the Geometry of Phenotype Space}},\ }\href
  {https://doi.org/10.1126/science.1217405} {\bibfield  {journal} {\bibinfo
  {journal} {Science}\ }\textbf {\bibinfo {volume} {336}},\ \bibinfo {pages}
  {1157} (\bibinfo {year} {2012})}\BibitemShut {NoStop}%
\bibitem [{\citenamefont {Seoane}\ and\ \citenamefont
  {Sol{\ifmmode\acute{e}\else\'{e}\fi}}(2016)}]{Seoane2016May}%
  \BibitemOpen
  \bibfield  {author} {\bibinfo {author} {\bibfnamefont {L.~F.}\ \bibnamefont
  {Seoane}}\ and\ \bibinfo {author} {\bibfnamefont {R.}~\bibnamefont
  {Sol{\ifmmode\acute{e}\else\'{e}\fi}}},\ }\bibfield  {title} {\bibinfo
  {title} {{Multiobjective Optimization and Phase Transitions}},\ }in\ \href
  {https://doi.org/10.1007/978-3-319-29228-1_22} {\emph {\bibinfo {booktitle}
  {{Proceedings of ECCS 2014}}}}\ (\bibinfo  {publisher} {Springer},\ \bibinfo
  {address} {Cham, Switzerland},\ \bibinfo {year} {2016})\ pp.\ \bibinfo
  {pages} {259--270}\BibitemShut {NoStop}%
\bibitem [{\citenamefont {Dinis}\ \emph {et~al.}(2020)\citenamefont {Dinis},
  \citenamefont {Unterberger},\ and\ \citenamefont {Lacoste}}]{Dinis2020Nov}%
  \BibitemOpen
  \bibfield  {author} {\bibinfo {author} {\bibfnamefont {L.}~\bibnamefont
  {Dinis}}, \bibinfo {author} {\bibfnamefont {J.}~\bibnamefont {Unterberger}},\
  and\ \bibinfo {author} {\bibfnamefont {D.}~\bibnamefont {Lacoste}},\
  }\bibfield  {title} {\bibinfo {title} {{Phase transitions in optimal betting
  strategies}},\ }\href {https://doi.org/10.1209/0295-5075/131/60005}
  {\bibfield  {journal} {\bibinfo  {journal} {Europhys. Lett.}\ }\textbf
  {\bibinfo {volume} {131}},\ \bibinfo {pages} {60005} (\bibinfo {year}
  {2020})}\BibitemShut {NoStop}%
\bibitem [{\citenamefont {Barato}\ and\ \citenamefont
  {Seifert}(2015{\natexlab{a}})}]{Barato2015Apr}%
  \BibitemOpen
  \bibfield  {author} {\bibinfo {author} {\bibfnamefont {A.~C.}\ \bibnamefont
  {Barato}}\ and\ \bibinfo {author} {\bibfnamefont {U.}~\bibnamefont
  {Seifert}},\ }\bibfield  {title} {\bibinfo {title} {{Thermodynamic
  Uncertainty Relation for Biomolecular Processes}},\ }\href
  {https://doi.org/10.1103/PhysRevLett.114.158101} {\bibfield  {journal}
  {\bibinfo  {journal} {Phys. Rev. Lett.}\ }\textbf {\bibinfo {volume} {114}},\
  \bibinfo {pages} {158101} (\bibinfo {year} {2015}{\natexlab{a}})}\BibitemShut
  {NoStop}%
\bibitem [{\citenamefont {Barato}\ and\ \citenamefont
  {Seifert}(2015{\natexlab{b}})}]{Barato2015Jun}%
  \BibitemOpen
  \bibfield  {author} {\bibinfo {author} {\bibfnamefont {A.~C.}\ \bibnamefont
  {Barato}}\ and\ \bibinfo {author} {\bibfnamefont {U.}~\bibnamefont
  {Seifert}},\ }\bibfield  {title} {\bibinfo {title} {{Universal Bound on the
  Fano Factor in Enzyme Kinetics}},\ }\href
  {https://doi.org/10.1021/acs.jpcb.5b01918} {\bibfield  {journal} {\bibinfo
  {journal} {J. Phys. Chem. B}\ }\textbf {\bibinfo {volume} {119}},\ \bibinfo
  {pages} {6555} (\bibinfo {year} {2015}{\natexlab{b}})}\BibitemShut {NoStop}%
\bibitem [{\citenamefont {Seifert}(2019)}]{Seifert2019Mar}%
  \BibitemOpen
  \bibfield  {author} {\bibinfo {author} {\bibfnamefont {U.}~\bibnamefont
  {Seifert}},\ }\bibfield  {title} {\bibinfo {title} {{From Stochastic
  Thermodynamics to Thermodynamic Inference}},\ }\href
  {https://doi.org/10.1146/annurev-conmatphys-031218-013554} {\bibfield
  {journal} {\bibinfo  {journal} {Annu. Rev. Condens. Matter Phys.}\ ,\
  \bibinfo {pages} {171}} (\bibinfo {year} {2019})}\BibitemShut {NoStop}%
\bibitem [{\citenamefont {Gingrich}\ \emph {et~al.}(2016)\citenamefont
  {Gingrich}, \citenamefont {Horowitz}, \citenamefont {Perunov},\ and\
  \citenamefont {England}}]{England}%
  \BibitemOpen
  \bibfield  {author} {\bibinfo {author} {\bibfnamefont {T.~R.}\ \bibnamefont
  {Gingrich}}, \bibinfo {author} {\bibfnamefont {J.~M.}\ \bibnamefont
  {Horowitz}}, \bibinfo {author} {\bibfnamefont {N.}~\bibnamefont {Perunov}},\
  and\ \bibinfo {author} {\bibfnamefont {J.~L.}\ \bibnamefont {England}},\
  }\bibfield  {title} {\bibinfo {title} {Dissipation bounds all steady-state
  current fluctuations},\ }\href
  {https://doi.org/10.1103/PhysRevLett.116.120601} {\bibfield  {journal}
  {\bibinfo  {journal} {Phys. Rev. Lett.}\ }\textbf {\bibinfo {volume} {116}},\
  \bibinfo {pages} {120601} (\bibinfo {year} {2016})}\BibitemShut {NoStop}%
\bibitem [{\citenamefont {Horowitz}\ and\ \citenamefont
  {Gingrich}(2019)}]{Hororwitz_2019}%
  \BibitemOpen
  \bibfield  {author} {\bibinfo {author} {\bibfnamefont {J.~M.}\ \bibnamefont
  {Horowitz}}\ and\ \bibinfo {author} {\bibfnamefont {T.~R.}\ \bibnamefont
  {Gingrich}},\ }\bibfield  {title} {\bibinfo {title} {Thermodynamic
  uncertainty relations constrain non-equilibrium fluctuations},\ }\href
  {https://doi.org/10.1038/s41567-019-0702-6} {\bibfield  {journal} {\bibinfo
  {journal} {Nat. Phys.}\ }\textbf {\bibinfo {volume} {16}},\ \bibinfo {pages}
  {15–20} (\bibinfo {year} {2019})}\BibitemShut {NoStop}%
\bibitem [{\citenamefont {Seifert}(2018)}]{Udo_a}%
  \BibitemOpen
  \bibfield  {author} {\bibinfo {author} {\bibfnamefont {U.}~\bibnamefont
  {Seifert}},\ }\bibfield  {title} {\bibinfo {title} {Stochastic
  thermodynamics: From principles to the cost of precision},\ }\href
  {https://doi.org/https://doi.org/10.1016/j.physa.2017.10.024} {\bibfield
  {journal} {\bibinfo  {journal} {Physica A}\ }\textbf {\bibinfo {volume}
  {504}},\ \bibinfo {pages} {176} (\bibinfo {year} {2018})},\ \bibinfo {note}
  {lecture Notes of the 14th International Summer School on Fundamental
  Problems in Statistical Physics}\BibitemShut {NoStop}%
\bibitem [{\citenamefont {Hartich}\ and\ \citenamefont
  {Godec}(2021)}]{Hartich}%
  \BibitemOpen
  \bibfield  {author} {\bibinfo {author} {\bibfnamefont {D.}~\bibnamefont
  {Hartich}}\ and\ \bibinfo {author} {\bibfnamefont {A.}~\bibnamefont
  {Godec}},\ }\bibfield  {title} {\bibinfo {title} {Thermodynamic uncertainty
  relation bounds the extent of anomalous diffusion},\ }\href
  {https://doi.org/10.1103/PhysRevLett.127.080601} {\bibfield  {journal}
  {\bibinfo  {journal} {Phys. Rev. Lett.}\ }\textbf {\bibinfo {volume} {127}},\
  \bibinfo {pages} {080601} (\bibinfo {year} {2021})}\BibitemShut {NoStop}%
\bibitem [{\citenamefont {Dieball}\ and\ \citenamefont
  {Godec}(2023)}]{Dieball}%
  \BibitemOpen
  \bibfield  {author} {\bibinfo {author} {\bibfnamefont {C.}~\bibnamefont
  {Dieball}}\ and\ \bibinfo {author} {\bibfnamefont {A.}~\bibnamefont
  {Godec}},\ }\bibfield  {title} {\bibinfo {title} {Direct route to
  thermodynamic uncertainty relations and their saturation},\ }\href
  {https://doi.org/10.1103/PhysRevLett.130.087101} {\bibfield  {journal}
  {\bibinfo  {journal} {Phys. Rev. Lett.}\ }\textbf {\bibinfo {volume} {130}},\
  \bibinfo {pages} {087101} (\bibinfo {year} {2023})}\BibitemShut {NoStop}%
\bibitem [{\citenamefont {Proesmans}(2023)}]{Proesmans}%
  \BibitemOpen
  \bibfield  {author} {\bibinfo {author} {\bibfnamefont {K.}~\bibnamefont
  {Proesmans}},\ }\bibfield  {title} {\bibinfo {title} {Precision-dissipation
  trade-off for driven stochastic systems},\ }\bibfield  {journal} {\bibinfo
  {journal} {Commun. Phys.}\ }\textbf {\bibinfo {volume} {6}},\ \href
  {https://doi.org/10.1038/s42005-023-01343-5} {10.1038/s42005-023-01343-5}
  (\bibinfo {year} {2023})\BibitemShut {NoStop}%
\bibitem [{\citenamefont {Lan}\ \emph {et~al.}(2012)\citenamefont {Lan},
  \citenamefont {Sartori}, \citenamefont {Neumann}, \citenamefont {Sourjik},\
  and\ \citenamefont {Tu}}]{Lan2012May}%
  \BibitemOpen
  \bibfield  {author} {\bibinfo {author} {\bibfnamefont {G.}~\bibnamefont
  {Lan}}, \bibinfo {author} {\bibfnamefont {P.}~\bibnamefont {Sartori}},
  \bibinfo {author} {\bibfnamefont {S.}~\bibnamefont {Neumann}}, \bibinfo
  {author} {\bibfnamefont {V.}~\bibnamefont {Sourjik}},\ and\ \bibinfo {author}
  {\bibfnamefont {Y.}~\bibnamefont {Tu}},\ }\bibfield  {title} {\bibinfo
  {title} {{The energy{\textendash}speed{\textendash}accuracy trade-off in
  sensory adaptation}},\ }\href {https://doi.org/10.1038/nphys2276} {\bibfield
  {journal} {\bibinfo  {journal} {Nat. Phys.}\ }\textbf {\bibinfo {volume}
  {8}},\ \bibinfo {pages} {422} (\bibinfo {year} {2012})}\BibitemShut {NoStop}%
\bibitem [{\citenamefont {Mehra}\ \emph {et~al.}(2006)\citenamefont {Mehra},
  \citenamefont {Hong}, \citenamefont {Shi}, \citenamefont {Loros},
  \citenamefont {Dunlap},\ and\ \citenamefont {Ruoff}}]{Mehra2006Jul}%
  \BibitemOpen
  \bibfield  {author} {\bibinfo {author} {\bibfnamefont {A.}~\bibnamefont
  {Mehra}}, \bibinfo {author} {\bibfnamefont {C.~I.}\ \bibnamefont {Hong}},
  \bibinfo {author} {\bibfnamefont {M.}~\bibnamefont {Shi}}, \bibinfo {author}
  {\bibfnamefont {J.~J.}\ \bibnamefont {Loros}}, \bibinfo {author}
  {\bibfnamefont {J.~C.}\ \bibnamefont {Dunlap}},\ and\ \bibinfo {author}
  {\bibfnamefont {P.}~\bibnamefont {Ruoff}},\ }\bibfield  {title} {\bibinfo
  {title} {{Circadian Rhythmicity by Autocatalysis}},\ }\href
  {https://doi.org/10.1371/journal.pcbi.0020096} {\bibfield  {journal}
  {\bibinfo  {journal} {PLoS Comput. Biol.}\ }\textbf {\bibinfo {volume} {2}},\
  \bibinfo {pages} {e96} (\bibinfo {year} {2006})}\BibitemShut {NoStop}%
\bibitem [{\citenamefont {Dobroborsky}(2006)}]{dobroborsky2006thermodynamics}%
  \BibitemOpen
  \bibfield  {author} {\bibinfo {author} {\bibfnamefont {B.}~\bibnamefont
  {Dobroborsky}},\ }\bibfield  {title} {\bibinfo {title} {Thermodynamics of
  biological systems},\ }\href@noop {} {\bibfield  {journal} {\bibinfo
  {journal} {North-Western State Medical University Press: Saint-Petersburg,
  Russia}\ } (\bibinfo {year} {2006})}\BibitemShut {NoStop}%
\bibitem [{\citenamefont {Qian}\ and\ \citenamefont
  {Reluga}(2005)}]{Qian2005Jan}%
  \BibitemOpen
  \bibfield  {author} {\bibinfo {author} {\bibfnamefont {H.}~\bibnamefont
  {Qian}}\ and\ \bibinfo {author} {\bibfnamefont {T.~C.}\ \bibnamefont
  {Reluga}},\ }\bibfield  {title} {\bibinfo {title} {{Nonequilibrium
  Thermodynamics and Nonlinear Kinetics in a Cellular Signaling Switch}},\
  }\href {https://doi.org/10.1103/PhysRevLett.94.028101} {\bibfield  {journal}
  {\bibinfo  {journal} {Phys. Rev. Lett.}\ }\textbf {\bibinfo {volume} {94}},\
  \bibinfo {pages} {028101} (\bibinfo {year} {2005})}\BibitemShut {NoStop}%
\bibitem [{\citenamefont {Goldbeter}\ and\ \citenamefont
  {Koshland}(1981)}]{Goldbeter1981Nov}%
  \BibitemOpen
  \bibfield  {author} {\bibinfo {author} {\bibfnamefont {A.}~\bibnamefont
  {Goldbeter}}\ and\ \bibinfo {author} {\bibfnamefont {D.~E.}\ \bibnamefont
  {Koshland}},\ }\bibfield  {title} {\bibinfo {title} {{An amplified
  sensitivity arising from covalent modification in biological systems.}},\
  }\href {https://doi.org/10.1073/pnas.78.11.6840} {\bibfield  {journal}
  {\bibinfo  {journal} {Proc. Natl. Acad. Sci. U.S.A.}\ }\textbf {\bibinfo
  {volume} {78}},\ \bibinfo {pages} {6840} (\bibinfo {year}
  {1981})}\BibitemShut {NoStop}%
\bibitem [{\citenamefont {Koshland}\ \emph {et~al.}(1982)\citenamefont
  {Koshland}, \citenamefont {Goldbeter},\ and\ \citenamefont
  {Stock}}]{Koshland1982Jul}%
  \BibitemOpen
  \bibfield  {author} {\bibinfo {author} {\bibfnamefont {D.~E.}\ \bibnamefont
  {Koshland}}, \bibinfo {author} {\bibfnamefont {A.}~\bibnamefont
  {Goldbeter}},\ and\ \bibinfo {author} {\bibfnamefont {J.~B.}\ \bibnamefont
  {Stock}},\ }\bibfield  {title} {\bibinfo {title} {{Amplification and
  Adaptation in Regulatory and Sensory Systems}},\ }\href
  {https://doi.org/10.1126/science.7089556} {\bibfield  {journal} {\bibinfo
  {journal} {Science}\ }\textbf {\bibinfo {volume} {217}},\ \bibinfo {pages}
  {220} (\bibinfo {year} {1982})}\BibitemShut {NoStop}%
\bibitem [{\citenamefont {Pigllucci}(1996)}]{pigllucci1996organisms}%
  \BibitemOpen
  \bibfield  {author} {\bibinfo {author} {\bibfnamefont {M.}~\bibnamefont
  {Pigllucci}},\ }\bibfield  {title} {\bibinfo {title} {How organisms respond
  to environmental changes: from phenotypes to molecules (and vice versa)},\
  }\href@noop {} {\bibfield  {journal} {\bibinfo  {journal} {Trends in Ecology
  \& Evolution}\ }\textbf {\bibinfo {volume} {11}},\ \bibinfo {pages} {168}
  (\bibinfo {year} {1996})}\BibitemShut {NoStop}%
\bibitem [{\citenamefont {Marchler}\ and\ \citenamefont
  {Wu}(2001)}]{Marchler2001Feb}%
  \BibitemOpen
  \bibfield  {author} {\bibinfo {author} {\bibfnamefont {G.}~\bibnamefont
  {Marchler}}\ and\ \bibinfo {author} {\bibfnamefont {C.}~\bibnamefont {Wu}},\
  }\bibfield  {title} {\bibinfo {title} {{Modulation of Drosophila heat shock
  transcription factor activity by the molecular chaperone DROJ1}},\ }\bibfield
   {journal} {\bibinfo  {journal} {EMBO J.}\ }\href
  {https://doi.org/10.1093/emboj/20.3.499} {10.1093/emboj/20.3.499} (\bibinfo
  {year} {2001})\BibitemShut {NoStop}%
\bibitem [{\citenamefont {Feldman}\ and\ \citenamefont
  {Frydman}(2000)}]{Feldman2000Feb}%
  \BibitemOpen
  \bibfield  {author} {\bibinfo {author} {\bibfnamefont {D.~E.}\ \bibnamefont
  {Feldman}}\ and\ \bibinfo {author} {\bibfnamefont {J.}~\bibnamefont
  {Frydman}},\ }\bibfield  {title} {\bibinfo {title} {{Protein folding in vivo:
  the importance of molecular chaperones}},\ }\href
  {https://doi.org/10.1016/S0959-440X(99)00044-5} {\bibfield  {journal}
  {\bibinfo  {journal} {Curr. Opin. Struct. Biol.}\ }\textbf {\bibinfo {volume}
  {10}},\ \bibinfo {pages} {26} (\bibinfo {year} {2000})}\BibitemShut {NoStop}%
\bibitem [{\citenamefont {Munakata}\ and\ \citenamefont
  {Rosinberg}(2013)}]{Munakata2013Jun}%
  \BibitemOpen
  \bibfield  {author} {\bibinfo {author} {\bibfnamefont {T.}~\bibnamefont
  {Munakata}}\ and\ \bibinfo {author} {\bibfnamefont {M.~L.}\ \bibnamefont
  {Rosinberg}},\ }\bibfield  {title} {\bibinfo {title} {{Feedback cooling,
  measurement errors, and entropy production}},\ }\href
  {https://doi.org/10.1088/1742-5468/2013/06/P06014} {\bibfield  {journal}
  {\bibinfo  {journal} {J. Stat. Mech.: Theory Exp.}\ }\textbf {\bibinfo
  {volume} {2013}}\bibinfo  {number} { (06)},\ \bibinfo {pages}
  {P06014}}\BibitemShut {NoStop}%
\bibitem [{\citenamefont {Munakata}\ and\ \citenamefont
  {Rosinberg}(2014)}]{Munakata2014May}%
  \BibitemOpen
\bibfield  {number} {  }\bibfield  {author} {\bibinfo {author} {\bibfnamefont
  {T.}~\bibnamefont {Munakata}}\ and\ \bibinfo {author} {\bibfnamefont {M.~L.}\
  \bibnamefont {Rosinberg}},\ }\bibfield  {title} {\bibinfo {title} {{Entropy
  Production and Fluctuation Theorems for Langevin Processes under Continuous
  Non-Markovian Feedback Control}},\ }\href
  {https://doi.org/10.1103/PhysRevLett.112.180601} {\bibfield  {journal}
  {\bibinfo  {journal} {Phys. Rev. Lett.}\ }\textbf {\bibinfo {volume} {112}},\
  \bibinfo {pages} {180601} (\bibinfo {year} {2014})}\BibitemShut {NoStop}%
\bibitem [{\citenamefont {Horowitz}\ and\ \citenamefont
  {Sandberg}(2014)}]{Horowitz2014Dec}%
  \BibitemOpen
  \bibfield  {author} {\bibinfo {author} {\bibfnamefont {J.~M.}\ \bibnamefont
  {Horowitz}}\ and\ \bibinfo {author} {\bibfnamefont {H.}~\bibnamefont
  {Sandberg}},\ }\bibfield  {title} {\bibinfo {title} {{Second-law-like
  inequalities with information and their interpretations}},\ }\href
  {https://doi.org/10.1088/1367-2630/16/12/125007} {\bibfield  {journal}
  {\bibinfo  {journal} {New J. Phys.}\ }\textbf {\bibinfo {volume} {16}},\
  \bibinfo {pages} {125007} (\bibinfo {year} {2014})}\BibitemShut {NoStop}%
\bibitem [{\citenamefont {Tu}\ \emph {et~al.}(2008)\citenamefont {Tu},
  \citenamefont {Shimizu},\ and\ \citenamefont {Berg}}]{Tu2008Sep}%
  \BibitemOpen
  \bibfield  {author} {\bibinfo {author} {\bibfnamefont {Y.}~\bibnamefont
  {Tu}}, \bibinfo {author} {\bibfnamefont {T.~S.}\ \bibnamefont {Shimizu}},\
  and\ \bibinfo {author} {\bibfnamefont {H.~C.}\ \bibnamefont {Berg}},\
  }\bibfield  {title} {\bibinfo {title} {{Modeling the chemotactic response of
  Escherichia coli to time-varying stimuli}},\ }\href
  {https://doi.org/10.1073/pnas.0807569105} {\bibfield  {journal} {\bibinfo
  {journal} {Proc. Natl. Acad. Sci. U.S.A.}\ }\textbf {\bibinfo {volume}
  {105}},\ \bibinfo {pages} {14855} (\bibinfo {year} {2008})}\BibitemShut
  {NoStop}%
\bibitem [{\citenamefont {Sartori}\ \emph {et~al.}(2014)\citenamefont
  {Sartori}, \citenamefont {Granger}, \citenamefont {Lee},\ and\ \citenamefont
  {Horowitz}}]{Sartori2014Dec}%
  \BibitemOpen
  \bibfield  {author} {\bibinfo {author} {\bibfnamefont {P.}~\bibnamefont
  {Sartori}}, \bibinfo {author} {\bibfnamefont {L.}~\bibnamefont {Granger}},
  \bibinfo {author} {\bibfnamefont {C.~F.}\ \bibnamefont {Lee}},\ and\ \bibinfo
  {author} {\bibfnamefont {J.~M.}\ \bibnamefont {Horowitz}},\ }\bibfield
  {title} {\bibinfo {title} {{Thermodynamic Costs of Information Processing in
  Sensory Adaptation}},\ }\href {https://doi.org/10.1371/journal.pcbi.1003974}
  {\bibfield  {journal} {\bibinfo  {journal} {PLoS Comput. Biol.}\ }\textbf
  {\bibinfo {volume} {10}},\ \bibinfo {pages} {e1003974} (\bibinfo {year}
  {2014})}\BibitemShut {NoStop}%
\bibitem [{\citenamefont {Conti}\ and\ \citenamefont
  {Mora}(2022)}]{Conti2022Nov}%
  \BibitemOpen
  \bibfield  {author} {\bibinfo {author} {\bibfnamefont {D.}~\bibnamefont
  {Conti}}\ and\ \bibinfo {author} {\bibfnamefont {T.}~\bibnamefont {Mora}},\
  }\bibfield  {title} {\bibinfo {title} {{Nonequilibrium dynamics of adaptation
  in sensory systems}},\ }\href {https://doi.org/10.1103/PhysRevE.106.054404}
  {\bibfield  {journal} {\bibinfo  {journal} {Phys. Rev. E}\ }\textbf {\bibinfo
  {volume} {106}},\ \bibinfo {pages} {054404} (\bibinfo {year}
  {2022})}\BibitemShut {NoStop}%
\bibitem [{\citenamefont {Van~den Broeck}\ and\ \citenamefont
  {Esposito}(2010)}]{VandenBroeck2010Jul}%
  \BibitemOpen
  \bibfield  {author} {\bibinfo {author} {\bibfnamefont {C.}~\bibnamefont
  {Van~den Broeck}}\ and\ \bibinfo {author} {\bibfnamefont {M.}~\bibnamefont
  {Esposito}},\ }\bibfield  {title} {\bibinfo {title} {{Three faces of the
  second law. II. Fokker-Planck formulation}},\ }\href
  {https://doi.org/10.1103/PhysRevE.82.011144} {\bibfield  {journal} {\bibinfo
  {journal} {Phys. Rev. E}\ }\textbf {\bibinfo {volume} {82}},\ \bibinfo
  {pages} {011144} (\bibinfo {year} {2010})}\BibitemShut {NoStop}%
\bibitem [{\citenamefont {Dieball}\ \emph {et~al.}(2023)\citenamefont
  {Dieball}, \citenamefont {Wellecke},\ and\ \citenamefont
  {Godec}}]{Diebal_asymm}%
  \BibitemOpen
  \bibfield  {author} {\bibinfo {author} {\bibfnamefont {C.}~\bibnamefont
  {Dieball}}, \bibinfo {author} {\bibfnamefont {G.}~\bibnamefont {Wellecke}},\
  and\ \bibinfo {author} {\bibfnamefont {A.}~\bibnamefont {Godec}},\ }\bibfield
   {title} {\bibinfo {title} {Asymmetric thermal relaxation in driven systems:
  Rotations go opposite ways},\ }\href
  {https://doi.org/10.1103/PhysRevResearch.5.L042030} {\bibfield  {journal}
  {\bibinfo  {journal} {Phys. Rev. Res.}\ }\textbf {\bibinfo {volume} {5}},\
  \bibinfo {pages} {L042030} (\bibinfo {year} {2023})}\BibitemShut {NoStop}%
\bibitem [{\citenamefont {Ibáñez}\ \emph {et~al.}(2024)\citenamefont
  {Ibáñez}, \citenamefont {Dieball}, \citenamefont {Lasanta}, \citenamefont
  {Godec},\ and\ \citenamefont {Rica}}]{Ibanez_2024}%
  \BibitemOpen
  \bibfield  {author} {\bibinfo {author} {\bibfnamefont {M.}~\bibnamefont
  {Ibáñez}}, \bibinfo {author} {\bibfnamefont {C.}~\bibnamefont {Dieball}},
  \bibinfo {author} {\bibfnamefont {A.}~\bibnamefont {Lasanta}}, \bibinfo
  {author} {\bibfnamefont {A.}~\bibnamefont {Godec}},\ and\ \bibinfo {author}
  {\bibfnamefont {R.~A.}\ \bibnamefont {Rica}},\ }\bibfield  {title} {\bibinfo
  {title} {Heating and cooling are fundamentally asymmetric and evolve along
  distinct pathways},\ }\href {https://doi.org/10.1038/s41567-023-02269-z}
  {\bibfield  {journal} {\bibinfo  {journal} {Nat. Phys.}\ }\textbf {\bibinfo
  {volume} {20}},\ \bibinfo {pages} {135–141} (\bibinfo {year}
  {2024})}\BibitemShut {NoStop}%
\bibitem [{\citenamefont {Dieball}\ and\ \citenamefont
  {Godec}(2024)}]{Cai_transport}%
  \BibitemOpen
  \bibfield  {author} {\bibinfo {author} {\bibfnamefont {C.}~\bibnamefont
  {Dieball}}\ and\ \bibinfo {author} {\bibfnamefont {A.}~\bibnamefont
  {Godec}},\ }\bibfield  {title} {\bibinfo {title} {Thermodynamic bounds on
  generalized transport: From single-molecule to bulk observables},\ }\href
  {https://doi.org/10.1103/PhysRevLett.133.067101} {\bibfield  {journal}
  {\bibinfo  {journal} {Phys. Rev. Lett.}\ }\textbf {\bibinfo {volume} {133}},\
  \bibinfo {pages} {067101} (\bibinfo {year} {2024})}\BibitemShut {NoStop}%
\bibitem [{\citenamefont {Goda}\ \emph {et~al.}(2014)\citenamefont {Goda},
  \citenamefont {Sharp},\ and\ \citenamefont {Wijnen}}]{Goda2014Oct}%
  \BibitemOpen
  \bibfield  {author} {\bibinfo {author} {\bibfnamefont {T.}~\bibnamefont
  {Goda}}, \bibinfo {author} {\bibfnamefont {B.}~\bibnamefont {Sharp}},\ and\
  \bibinfo {author} {\bibfnamefont {H.}~\bibnamefont {Wijnen}},\ }\bibfield
  {title} {\bibinfo {title} {{Temperature-dependent resetting of the molecular
  circadian oscillator in Drosophila}},\ }\bibfield  {journal} {\bibinfo
  {journal} {Proc. R. Soc. B.}\ }\textbf {\bibinfo {volume} {281}},\ \href
  {https://doi.org/10.1098/rspb.2014.1714} {10.1098/rspb.2014.1714} (\bibinfo
  {year} {2014})\BibitemShut {NoStop}%
\bibitem [{\citenamefont {Mori}\ \emph {et~al.}(2023)\citenamefont {Mori},
  \citenamefont {Olsen},\ and\ \citenamefont {Krishnamurthy}}]{Mori2023May}%
  \BibitemOpen
  \bibfield  {author} {\bibinfo {author} {\bibfnamefont {F.}~\bibnamefont
  {Mori}}, \bibinfo {author} {\bibfnamefont {K.~S.}\ \bibnamefont {Olsen}},\
  and\ \bibinfo {author} {\bibfnamefont {S.}~\bibnamefont {Krishnamurthy}},\
  }\bibfield  {title} {\bibinfo {title} {{Entropy production of resetting
  processes}},\ }\href {https://doi.org/10.1103/PhysRevResearch.5.023103}
  {\bibfield  {journal} {\bibinfo  {journal} {Phys. Rev. Res.}\ }\textbf
  {\bibinfo {volume} {5}},\ \bibinfo {pages} {023103} (\bibinfo {year}
  {2023})}\BibitemShut {NoStop}%
\bibitem [{\citenamefont {Esposito}\ and\ \citenamefont {Van~den
  Broeck}(2010)}]{Esposito2010Jul}%
  \BibitemOpen
  \bibfield  {author} {\bibinfo {author} {\bibfnamefont {M.}~\bibnamefont
  {Esposito}}\ and\ \bibinfo {author} {\bibfnamefont {C.}~\bibnamefont {Van~den
  Broeck}},\ }\bibfield  {title} {\bibinfo {title} {{Three faces of the second
  law. I. Master equation formulation}},\ }\href
  {https://doi.org/10.1103/PhysRevE.82.011143} {\bibfield  {journal} {\bibinfo
  {journal} {Phys. Rev. E}\ }\textbf {\bibinfo {volume} {82}},\ \bibinfo
  {pages} {011143} (\bibinfo {year} {2010})}\BibitemShut {NoStop}%
\bibitem [{\citenamefont {Depken}\ \emph {et~al.}(2013)\citenamefont {Depken},
  \citenamefont {Parrondo},\ and\ \citenamefont {Grill}}]{Depken2013Oct}%
  \BibitemOpen
  \bibfield  {author} {\bibinfo {author} {\bibfnamefont {M.}~\bibnamefont
  {Depken}}, \bibinfo {author} {\bibfnamefont {J.~M.~R.}\ \bibnamefont
  {Parrondo}},\ and\ \bibinfo {author} {\bibfnamefont {S.~W.}\ \bibnamefont
  {Grill}},\ }\bibfield  {title} {\bibinfo {title} {{Intermittent Transcription
  Dynamics for the Rapid Production of Long Transcripts of High Fidelity}},\
  }\href {https://doi.org/10.1016/j.celrep.2013.09.007} {\bibfield  {journal}
  {\bibinfo  {journal} {Cell Rep.}\ }\textbf {\bibinfo {volume} {5}},\ \bibinfo
  {pages} {521} (\bibinfo {year} {2013})}\BibitemShut {NoStop}%
\bibitem [{\citenamefont {Phillips}(2020)}]{Phillips2020Sep}%
  \BibitemOpen
  \bibfield  {author} {\bibinfo {author} {\bibfnamefont {R.}~\bibnamefont
  {Phillips}},\ }\href {https://doi.org/10.1515/9780691200255} {\emph {\bibinfo
  {title} {{The Molecular Switch}}}}\ (\bibinfo  {publisher} {Princeton
  University Press},\ \bibinfo {address} {Princeton, NJ, USA},\ \bibinfo {year}
  {2020})\BibitemShut {NoStop}%
\bibitem [{\citenamefont {Scott}\ and\ \citenamefont
  {Hwa}(2023)}]{Scott2023May}%
  \BibitemOpen
  \bibfield  {author} {\bibinfo {author} {\bibfnamefont {M.}~\bibnamefont
  {Scott}}\ and\ \bibinfo {author} {\bibfnamefont {T.}~\bibnamefont {Hwa}},\
  }\bibfield  {title} {\bibinfo {title} {{Shaping bacterial gene expression by
  physiological and proteome allocation constraints}},\ }\href
  {https://doi.org/10.1038/s41579-022-00818-6} {\bibfield  {journal} {\bibinfo
  {journal} {Nat. Rev. Microbiol.}\ }\textbf {\bibinfo {volume} {21}},\
  \bibinfo {pages} {327} (\bibinfo {year} {2023})}\BibitemShut {NoStop}%
\bibitem [{\citenamefont {Cossetto}\ \emph {et~al.}(2024)\citenamefont
  {Cossetto}, \citenamefont {Rodenfels},\ and\ \citenamefont
  {Sartori}}]{Cossetto2024Aug}%
  \BibitemOpen
  \bibfield  {author} {\bibinfo {author} {\bibfnamefont {T.}~\bibnamefont
  {Cossetto}}, \bibinfo {author} {\bibfnamefont {J.}~\bibnamefont
  {Rodenfels}},\ and\ \bibinfo {author} {\bibfnamefont {P.}~\bibnamefont
  {Sartori}},\ }\bibfield  {title} {\bibinfo {title} {{Thermodynamic
  dissipation constrains metabolic versatility of unicellular growth}},\ }\href
  {https://doi.org/10.1101/2024.03.21.585772} {\bibfield  {journal} {\bibinfo
  {journal} {bioRxiv}\ ,\ \bibinfo {pages} {2024.03.21.585772}} (\bibinfo
  {year} {2024})},\ \Eprint {https://arxiv.org/abs/2024.03.21.585772}
  {2024.03.21.585772} \BibitemShut {NoStop}%
\bibitem [{\citenamefont {Calabrese}\ \emph {et~al.}(2021)\citenamefont
  {Calabrese}, \citenamefont {Chakrawal}, \citenamefont {Manzoni},\ and\
  \citenamefont {Van~Cappellen}}]{Calabrese2021Nov}%
  \BibitemOpen
  \bibfield  {author} {\bibinfo {author} {\bibfnamefont {S.}~\bibnamefont
  {Calabrese}}, \bibinfo {author} {\bibfnamefont {A.}~\bibnamefont
  {Chakrawal}}, \bibinfo {author} {\bibfnamefont {S.}~\bibnamefont {Manzoni}},\
  and\ \bibinfo {author} {\bibfnamefont {P.}~\bibnamefont {Van~Cappellen}},\
  }\bibfield  {title} {\bibinfo {title} {{Energetic scaling in microbial
  growth}},\ }\href {https://doi.org/10.1073/pnas.2107668118} {\bibfield
  {journal} {\bibinfo  {journal} {Proc. Natl. Acad. Sci. U.S.A.}\ }\textbf
  {\bibinfo {volume} {118}},\ \bibinfo {pages} {e2107668118} (\bibinfo {year}
  {2021})}\BibitemShut {NoStop}%
\bibitem [{\citenamefont {Bettencourt}\ \emph {et~al.}(2002)\citenamefont
  {Bettencourt}, \citenamefont {Kim}, \citenamefont {Hoffmann},\ and\
  \citenamefont {Feder}}]{Bettencourt2002Sep}%
  \BibitemOpen
  \bibfield  {author} {\bibinfo {author} {\bibfnamefont {B.~R.}\ \bibnamefont
  {Bettencourt}}, \bibinfo {author} {\bibfnamefont {I.}~\bibnamefont {Kim}},
  \bibinfo {author} {\bibfnamefont {A.~A.}\ \bibnamefont {Hoffmann}},\ and\
  \bibinfo {author} {\bibfnamefont {M.~E.}\ \bibnamefont {Feder}},\ }\bibfield
  {title} {\bibinfo {title} {{Response To Natural And Laboratory Selection At
  The Drosophila Hsp70 Genes}},\ }\href
  {https://doi.org/10.1111/j.0014-3820.2002.tb00193.x} {\bibfield  {journal}
  {\bibinfo  {journal} {Evolution}\ }\textbf {\bibinfo {volume} {56}},\
  \bibinfo {pages} {1796} (\bibinfo {year} {2002})}\BibitemShut {NoStop}%
\bibitem [{\citenamefont {Hoffmann}\ and\ \citenamefont
  {Harshman}(1999)}]{Hoffmann1999Dec}%
  \BibitemOpen
  \bibfield  {author} {\bibinfo {author} {\bibfnamefont {A.~A.}\ \bibnamefont
  {Hoffmann}}\ and\ \bibinfo {author} {\bibfnamefont {L.~G.}\ \bibnamefont
  {Harshman}},\ }\bibfield  {title} {\bibinfo {title} {{Desiccation and
  starvation resistance in Drosophila: patterns of variation at the species,
  population and intrapopulation levels}},\ }\href
  {https://doi.org/10.1046/j.1365-2540.1999.00649.x} {\bibfield  {journal}
  {\bibinfo  {journal} {Heredity}\ }\textbf {\bibinfo {volume} {83}},\ \bibinfo
  {pages} {637} (\bibinfo {year} {1999})}\BibitemShut {NoStop}%
\end{thebibliography}%

\end{document}